\documentclass[numbers]{aastex631}

	\usepackage{graphicx}	
	\usepackage{amsmath}
	\usepackage{amssymb}	
	\usepackage{longtable}
	\usepackage{threeparttable}
	\usepackage{multirow}
	\usepackage{booktabs}
	\usepackage{array}
    \usepackage[figuresright]{rotating}
    \usepackage[T1]{fontenc}

	\shorttitle{Cataclysmic Variable Stars with Negative Superhumps}
	\shortauthors{Sun et al.}
	\received{October 1, 2023}
	\accepted{November 21, 2023}

	\graphicspath{{./}{figures/}}

\begin{document}
			\title{Nine New Cataclysmic Variable Stars with Negative Superhumps}


\correspondingauthor{Sheng-Bang Qian, Qi-Bin Sun}
\email{qsb@ynao.ac.cn and sunqibin@ynao.ac.cn}	
\author{Qi-Bin Sun}
\affiliation{Yunnan Observatories, Chinese Academy of Sciences, Kunming 650216, PR of China}
\affiliation{Department of Astronomy, Key Laboratory of Astroparticle Physics of Yunnan Province, Yunnan University, Kunming 650091, PR of China}
\affiliation{Key Laboratory of the Structure and Evolution of Celestial Objects, Chinese Academy of Sciences, Kunming 650216, PR China}
\affiliation{Center for Astronomical Mega-Science, Chinese Academy of Sciences, 20A Datun Road, Chaoyang District, Beijing, 100012, PR of China}
\affiliation{University of Chinese Academy of Sciences, No.19(A) Yuquan Road, Shijingshan District, Beijing, PR of China}	

\author{Sheng-Bang Qian}

\affiliation{Yunnan Observatories, Chinese Academy of Sciences, Kunming 650216, PR of China}
\affiliation{Department of Astronomy, Key Laboratory of Astroparticle Physics of Yunnan Province, Yunnan University, Kunming 650091, PR of China}
\affiliation{Key Laboratory of the Structure and Evolution of Celestial Objects, Chinese Academy of Sciences, Kunming 650216, PR China}
\affiliation{Center for Astronomical Mega-Science, Chinese Academy of Sciences, 20A Datun Road, Chaoyang District, Beijing, 100012, PR of China}
\affiliation{University of Chinese Academy of Sciences, No.19(A) Yuquan Road, Shijingshan District, Beijing, PR of China}	

\author{Li-Ying Zhu}
\affiliation{Yunnan Observatories, Chinese Academy of Sciences, Kunming 650216, PR of China}
\affiliation{Key Laboratory of the Structure and Evolution of Celestial Objects, Chinese Academy of Sciences, Kunming 650216, PR China}
\affiliation{Center for Astronomical Mega-Science, Chinese Academy of Sciences, 20A Datun Road, Chaoyang District, Beijing, 100012, PR of China}
\affiliation{University of Chinese Academy of Sciences, No.19(A) Yuquan Road, Shijingshan District, Beijing, PR of China}

\author{Wen-Ping Liao}
\affiliation{Yunnan Observatories, Chinese Academy of Sciences, Kunming 650216, PR of China}
\affiliation{Key Laboratory of the Structure and Evolution of Celestial Objects, Chinese Academy of Sciences, Kunming 650216, PR China}
\affiliation{Center for Astronomical Mega-Science, Chinese Academy of Sciences, 20A Datun Road, Chaoyang District, Beijing, 100012, PR of China}
\affiliation{University of Chinese Academy of Sciences, No.19(A) Yuquan Road, Shijingshan District, Beijing, PR of China}

\author{Er-Gang Zhao}
\affiliation{Yunnan Observatories, Chinese Academy of Sciences, Kunming 650216, PR of China}
\affiliation{Key Laboratory of the Structure and Evolution of Celestial Objects, Chinese Academy of Sciences, Kunming 650216, PR China}
\affiliation{Center for Astronomical Mega-Science, Chinese Academy of Sciences, 20A Datun Road, Chaoyang District, Beijing, 100012, PR of China}

\author{Fu-Xing Li}
\affiliation{Yunnan Observatories, Chinese Academy of Sciences, Kunming 650216, PR of China}
\affiliation{Key Laboratory of the Structure and Evolution of Celestial Objects, Chinese Academy of Sciences, Kunming 650216, PR China}
\affiliation{Center for Astronomical Mega-Science, Chinese Academy of Sciences, 20A Datun Road, Chaoyang District, Beijing, 100012, PR of China}

\author{Xiang-Dong Shi}
\affiliation{Yunnan Observatories, Chinese Academy of Sciences, Kunming 650216, PR of China}
\affiliation{Key Laboratory of the Structure and Evolution of Celestial Objects, Chinese Academy of Sciences, Kunming 650216, PR China}
\affiliation{Center for Astronomical Mega-Science, Chinese Academy of Sciences, 20A Datun Road, Chaoyang District, Beijing, 100012, PR of China}

\author{Min-Yu Li}
\affiliation{Yunnan Observatories, Chinese Academy of Sciences, Kunming 650216, PR of China}
\affiliation{Key Laboratory of the Structure and Evolution of Celestial Objects, Chinese Academy of Sciences, Kunming 650216, PR China}
\affiliation{Center for Astronomical Mega-Science, Chinese Academy of Sciences, 20A Datun Road, Chaoyang District, Beijing, 100012, PR of China}
\affiliation{University of Chinese Academy of Sciences, No.19(A) Yuquan Road, Shijingshan District, Beijing, PR of China}			
			

			\begin{abstract}
Negative superhumps (NSHs) are signals a few percent shorter than the orbital period of a binary star and are considered to originate from the reverse precession of the tilted disk. Based on TESS photometry, we find nine new cataclysmic variable stars (CVs) with NSHs. Three (ASAS J1420, TZ Per, and V392 Hya) of these stars similar to AH Her still have NSHs during dwarf nova outbursts, and the NSH amplitude varies with the outburst. The variation in the radius of the accretion disk partially explains this phenomenon. However, it does not explain the rebound of the NSH amplitude after the peak of the outburst and the fact that the NSH amplitude of the quiescence is sometimes not the largest, and it is necessary to combine the disk instability model (DIM) and add other ingredients. Therefore, we suggest that the variation of NSH amplitude with outburst can be an essential basis for studying the origin of NSHs and improving the DIM. The six (ASASSN-V J1137, ASASSN-V J0611, 2MASS J0715, LAMOST J0925, ASASSN-17qj, and ZTF18acakuxo) remaining stars have been poorly studied, and for the first time we determine their orbital periods, NSHs and Superorbital signal (SOR) periods. The NSH periods and amplitudes of ASASSN-V J1137 and ASASSN-17qj vary with the SOR, and based on the comparison of the observations with the theory, we suggest that a single change in tilted disk angle does not explain the observations of the SOR and that other ingredients need to be considered as well.

\end{abstract}



\section{Introduction} \label{sec:intro}
Two types of hump modulations, positive superhumps (PSHs) and negative superhump (NSHs), have been found to exist in cataclysmic variable stars (CVs).
The period of the PSHs is a few percent longer than the orbital period and is thought to have originated from the line of apsides precession caused by the 3:1 tidal resonance between the accretion disk and the secondary star (e.g., \citealp{vogt1982z}; \citealp{osaki1985irradiation}; \citealp{wood2011v344}).
NSHs are another hump modulation a few percent shorter than the orbital period, are found in CVs and low-mass X-ray binaries (LMXBs) \citep{retter2002detection}. Since the discovery of the first NSHs system (TV Col; \citealp{Motch1981}), an increasing number of NSHs systems have been discovered (e.g., \citealp{Olech2009MNRAS}; \citealp{2013MNRAS.435..707A};  \citealp{sun2022study,Sun2023arXiv}; \citealp{Bruch2022,Bruch2023,BruchIII}). 
A large number of studies have suggested that the NSHs originate from the interaction between the reverse precession of the nodal line from the tilted disk and the orbital motion of the binary (e.g., \citealp{1985A&A...143..313B}; \citealp{patterson1999permanent}; \citealp{harvey1995superhumps}), but there is still no consensus on the exact physical process.
\cite{Barrett1988MNRAS} suggests that the NSHs may be caused by the change in kinetic energy of the gas stream as it impacts the surface of the tilted disk.
\cite{Patterson1997PASP} suggests that in scenarios where the accretion disk is tilted, the stream will easily collide into the inner disk, releasing more energy, and the NSHs arise from the beat between the tilted disk precession and the orbital motion.
\cite{Wood2007ApJ...661.1042W} suggests that the NSHs are caused by periodic changes in the position of the hot spot on the surface of the tilted disk, which causes the energy released by the stream to vary periodically. 

Evidence of tilted disk precession comes from the Superorbital signal (SOR). SOR is a periodic modulation with a period of about a few days and is thought to originate the tilted disk precession (e.g., \citealp{Katz1973NPhS..246...87K}; \citealp{Barrett1988MNRAS}; \citealp{harvey1995superhumps}), with orbital period and NSHs can be expressed as follows:
\begin{equation} \label{eq:pso}
	\dfrac{1}{P_{\rm prec}}  = \dfrac{1}{P_{\rm sor}} =\dfrac{1}{P_{\rm nsh}} - \dfrac{1}{P_{\rm orb}}  
\end{equation}
More evidence of accretion disk precession has been found recently.
\cite{Boyd2017} found that the eclipse depth, width, and skew of the nova-like variable star (NL) DW UMa vary with accretion disk precession, and they expressed the change in the slope of eclipse ingress and egress in terms of the change in eclipse skew.
\cite{2021MNRAS.503.4050I} studied the eclipse depths of AQ Men based on TESS photometry and also found that the eclipse depths of AQ Men showed a cyclic variation consistent with the accretion disk precession cycle.
In our recent study \citep{Sun2023arXiv}, we found that in NL SDSS J081256.85+191157.8, the eclipse depth, NSH amplitude, and O - C at the minimum of the eclipse, all have periodic variations similar to the accretion disk precession cycle, and all reach a maximum at the $ \sim $ 0.75 precession phases of the tilted disk (In this work we define the maximum value of the SOR signal as the zero phase of the tilted disk precession), providing new evidence that the NSHs are correlated with the tilted disk precession.
However, there are still many unanswered questions about the origin of the NSHs, such as the excitation mechanism of the tilted disk and reverse precession, how the material stream interacts with the accretion disk, why the amplitude of the NSH varies with the period of the tilted disk precession, and that the maximum of the eclipse depth does not occur at the maximum brightness of the observed, and so on. It is known that tilted disks, streams from secondary stars, hot spots, and binary star motions play a significant role in the origin of NSHs. Based on these four components, the NSHs have been reproduced using the smoothed particle hydrodynamic (SPH) simulations (e.g., \citealp{wood2009sph}; \citealp{Montgomery2009MNRAS}; \citealp{2021PASJ...73.1225K}).
Thus, the NSHs are an ideal window for the study of CVs and LMXBs.

Dwarf novae (DNe) are a subtype of CVs, consisting of a white dwarf (primary) and a late-type star main sequence star (secondary) \citep{warner1995cat}. The secondary material fills the Roche lobe, transferring material to the white dwarf, and the material from the secondary forms an accretion disk around the white dwarf. It is acknowledged that DNe can also be divided into three main subtypes: Z Cam, SU UMa, and U Gem. DNe have diverse outburst characteristics, with outburst amplitudes ranging from magnitudes 2 - 8 and durations from days to weeks, and outbursts are recurring, with recurrence times ranging from days to decades \citep{lasota2001disc}. Since the disk instability model (DIM) was proposed nearly 50 years ago \citep{1974PASJ...26..429O}, the DIM has been widely used to explain DN outbursts, and although it continues to face challenges, the DIM model, with the addition of other ingredients such as changes in the material transfer, inner disk truncation, accretion disk radiation, winds, and outflows etc. (e.g., \citealp{1994ApJ...427..956L}; \citealp{2015AcPPP...2..116B}; \citealp{2018Natur.554...69T}; \citealp{2020AdSpR..66.1004H}), continues to be used as the base model for explaining DN outbursts.

NSHs have been observed during both DN outbursts and quiescence (e.g., V503 Cyg,  \citealp{harvey1995superhumps}; ER UMa \citealp{Ohshima2012PASJ}; BK Lyn \citealp{Kato2013PASJ}; V1504 Cyg, \citealp{Ohshima2014PASJ}; V344 Lyr, \citealp{Wood2011ApJ}; SDSS J210014.12+004446.0, \citealp{Olech2009MNRAS}; NY Her, \citealp{Pavlenko2021Ap}).
However, \cite{ramsay2017v729} and \cite{court2020ex} at Z Cam-type DNe V729 Sgr and EX Dra both found NSHs only in the quiescence and were not found during the outbursts, demonstrating that there is no consensus on the existence of NSHs during DN outbursts. Our recent work found that the NSHs in the Z Cam-type DN AH Her changes as the outburst progresses \citep{2023arXiv230905891S}. Therefore, we suggest that the relationship between NSHs and DN outbursts is a window into the study of DN outbursts and the origin of NSH.
In the current study, we find that the NSHs are still present during the outbursts of three DNe (ASAS J1420, TZ Per, and V392 Hya) and that the NSH amplitude also varies with the outbursts, suggesting that the NSHs variation with outbursts in DNe may not be an exception but a general phenomenon.

In this paper, we discover nine new NSH systems based on TESS photometry, three of which have NSHs that vary with DN outbursts, and the other six are all found to have accretion disk precession signals and NSHs coexisting. This paper aims to provide new samples and phenomena for the study of the origin of NSHs and the improvement of the DNe outburst model. The paper is organized as follows.
Section \ref{sec:style} presents the \textit{TESS} photometry.
Section \ref{sec:Analysis} is the process and results of analyzing the nine stars.
Section \ref{sec:DISCUSSION} will discuss the relationship between the amplitudes of NSHs and DN outbursts and the variability of NSH and SOR, respectively.
Section \ref{sec:CONCLUSIONS} is the summary.

\begin{deluxetable*}{llllll}
\tablecaption{Journal of observations.\label{LOG}}
\tablewidth{0.01pt}
\tabletypesize{\scriptsize}
\tablehead{	
	\colhead{Stars}&\colhead{RA$^ {\rm a}$ }&\colhead{Dec$^ {\rm a}$}&\colhead{Sectors}&\colhead{ Start time$^ {\rm b}$ }&\colhead{End time$^ {\rm b}$ }\\
}
\startdata
ASAS J1420&14 20 23.47& -48 55 58.3&38&2333.85982&2360.55561\\
TZ Per             &02 13 50.96 &+58 22 52.3&58&2882.33382&2910.05198\\
V392 Hya           &10 58 56.42 &-29 14 40.8&36&2280.90841&2305.99315\\
&   &&63&3014.37341&3040.90384\\
ASASSN-V J1137&11 37 50.22 &-57 22 34.5&37&2308.50707&2332.58214\\
&&&64&3041.11789&3068.03454\\
ASASSN-V J0611&06 11 15.65 &+55 44 07.8&60&2939.47899&2962.58685\\
2MASS J0715&07 15 53.89 &+58 16 06.5&60&2939.90238&2962.58698\\    
LAMOST J0925&09 25 34.84& +43 49 17.8&21&1870.45053&1897.78503\\
ASASSN-17qj        &10 30 48.96 &-41 53 28.7&36&2282.18670&2305.99230\\
&&&63&3014.37252&3040.90302\\
ZTF18acakuxo       &04 29 28.67 &+60 40 32.9&59&2911.75915&2936.69099\\
\enddata
\begin{threeparttable} 
	\begin{tablenotes} 
		\item [a] Equatorial coordinates come from International Variable Star Index and are in the J2000 epoch.
\item [b] Start and end times are in BJD - 2457000.	
		
	\end{tablenotes} 
\end{threeparttable}
\end{deluxetable*}


\section{\textit{TESS} Photometry} \label{sec:style}

The \textit{Kepler Space Telescope K2 Missions} \citep{borucki2010kepler,howell2014k2} and the space telescope \textit{Transiting Exoplanet Survey Satellite} (\textit{TESS}, \citealp{Ricker2015journal}) have provided new opportunities for studies of CVs by providing high-resolution photometric data over an extended time base, and we have recently conducted a large number of studies based on data released by \textit{K2} and \textit{TESS}  (e.g., \citealp{2023arXiv230905891S}; \citealp{Sun2023arXiv}; \citealp{Sun2023MNRAS}; \citealp{sun2022study}; \citealp{2023ApJS..266...28L}). The K2's exposure time is up to 58 seconds, but the photometry is limited to the remaining part of the sky, while the \textit{TESS} photometry covers almost all of the sky. Moreover, \textit{K2} has been retired in 2018; in the present work, we use data from \textit{TESS}. The \textit{TESS Science Processing Operations Center} pipeline (SPOC; \citealp{Jenkins2016}) offers both Simple Aperture Photometry (SAP) and Pre-Search Data-Conditioned Simple Aperture Photometry (PDCSAP) light curves (for details, see \citealp{2010SPIE.7740E..1UT} and \citealp{2012PASP..124..963K}). PDCSAP attempts to reduce systematic errors but sometimes distorts DN outbursts. DN outburst profiles are easily distinguishable whether they are distorted or not, so in our study we preferred the PDCSAP data and used the SAP data only when DN outbursts were distorted. Among the DNe with outbursts in our study, only the PDCSAP fluxes for the star V392 Hya where the outbursts were distorted used SAP fluxes, and all studies of the other stars used PDCSAP fluxes. The sectors and spans of the \textit{TESS} photometry for the nine stars are shown in Table \ref{LOG}.

\begin{deluxetable*}{llllllllllllll}
\tablecaption{Frequency analysis results for Period04, where the meaning of the parameters is explained in Eq. \ref{eq:sine}.\label{data}}
\tablewidth{0.01pt}
\tabletypesize{\scriptsize}
\tablehead{	
	\colhead{ Stars and Sectors}&\colhead{Number}&\colhead{Types$ ^{\rm a} $}&\colhead{Z}&\colhead{ $\Phi_i$ }&\colhead{errors }&\colhead{P}&\colhead{ err}&\colhead{$A_i$}&\colhead{ errors}&\colhead{$\Omega_i$ }&\colhead{errors }&\colhead{Noise }&\colhead{S/N}\\
	\colhead{ }&\colhead{}&\colhead{ }&\colhead{[e/s]}&\colhead{[1/d]}&\colhead{[1/d]}&\colhead{[d]}&\colhead{ [d]}&\colhead{[e/s]}&\colhead{ [e/s]}&\colhead{[rad]] }&\colhead{[rad] }&\colhead{[e/s] }&\colhead{  }\\}
\startdata
ASAS J1420 (s38)&F1&orb&-0.085&3.80698 &0.00034 &0.262676 &0.000024 &10.320 &0.171 &0.510 &0.003 &1.492 &6.915 \\
&F2&nsh&&4.02165 &0.00026 &0.248654 &0.000016 &13.411 &0.171 &0.357 &0.002 &1.457 &9.202 \\
&F3&2*orb& &7.62070 &0.00025 &0.131222 &0.000004 &14.004 &0.171 &0.723 &0.002 &0.913 &15.343\\
&F4&3*orb&&11.43092 &0.00254 &0.087482 &0.000019 &1.390 &0.171 &0.587 &0.020 &0.230 &6.038 \\
\hline
TZ Per (s58) &F1&nsh&0.175 &4.01578 &0.00018 &0.249018 &0.000011 &26.421 &0.240 &0.690 &0.001 &2.452 &10.775 \\
&F2&2*orb&&7.60403 &0.00045 &0.131509 &0.000008 &10.558 &0.240 &0.574 &0.004 &1.232 &8.571 \\
\hline
V392 Hya (s36)&F1&orb&0.040 &3.07896 &0.00032 &0.324786 &0.000033 &2.852 &0.043 &0.891 &0.002 &0.510 &5.596 \\
&F2&nsh&&3.24838 &0.00016 &0.307846 &0.000015 &5.751 &0.043 &0.915 &0.001 &0.508 &11.332 \\
&F3&2*orb&&6.15377 &0.00026 &0.162502 &0.000007 &3.431 &0.043 &0.421 &0.002 &0.254 &13.510 \\
\hline
V392 Hya (s63)
&F1&orb&0.019&3.07393 &0.00066 &0.325316 &0.000070 &1.516 &0.046 &0.413 &0.005 &0.379 &4.004 \\
&F2&nsh& &3.24968 &0.00022 &0.307722 &0.000021 &4.595 &0.046 &0.539 &0.002 &0.397 &11.572 \\
&F3&2*orb&&6.15607 &0.00026 &0.162441 &0.000007 &3.789 &0.046 &0.421 &0.002 &0.272 &13.918\\
\hline
ASASSN-V J1137 (s37)&F1&sor&171.271 &0.26473 &0.00019 &3.777451 &0.002651 &26.333 &0.214 &0.370 &0.001 &3.815 &6.902\\
&F2&orb&&6.60552 &0.00061 &0.151389 &0.000014 &8.031 &0.214 &0.400 &0.004 &1.805 &4.449\\
&F3&nsh&&6.87288 &0.00034 &0.145499 &0.000007 &14.216 &0.214 &0.660 &0.002 &1.779 &7.990\\
\hline
ASASSN-V J1137 (s64)&F1&sor&170.232 &0.29392 &0.00022 &3.402234 &0.002512 &19.506 &0.207 &0.986 &0.002 &2.905 &6.716\\
&F2&orb&&6.60290 &0.00042 &0.151449 &0.000010 &10.037 &0.207 &0.417 &0.003 &1.594 &6.295\\
&F3&nsh&&6.89616 &0.00036 &0.145008 &0.000008 &11.831 &0.207 &0.384 &0.003 &1.541 &7.675\\
\hline
ASASSN-V J0611 (s60)&F1&sor&58.039 &0.29770 &0.00040 &3.359034 &0.004491 &6.244 &0.104 &0.030 &0.003 &1.166 &5.354\\
&F2&orb&&6.86474 &0.00085 &0.145672 &0.000018 &2.917 &0.104 &0.895 &0.006 &0.527 &5.535 \\
&F3&nsh&&7.16023 &0.00071 &0.139660 &0.000014 &3.477 &0.104 &0.551 &0.005 &0.526 &6.611 \\
\hline
2MASS J0715 (s60)&F1&sor&88.220 &0.32197 &0.00020 &3.105849 &0.001932 &11.079 &0.091 &0.003 &0.001 &1.812 &6.115\\
&F2&orb&&7.06724 &0.00079 &0.141498 &0.000016 &2.812 &0.091 &0.047 &0.005 &0.659 &4.267\\
&F3&nsh&&7.38915 &0.00040 &0.135333 &0.000007 &5.535 &0.091 &0.299 &0.003 &0.657 &8.423\\
\hline
LAMOST J0925 (s21)
&F1&sor&654.263&0.24765 &0.00036 &4.038024 &0.005931 &24.062 &0.434 &0.906 &0.003 &5.108 &4.711\\
&F2&orb&&6.74089 &0.00046 &0.148348 &0.000010 &19.091 &0.434 &0.579 &0.004 &3.145 &6.070\\
&F3&nsh& &6.98876 &0.00027 &0.143087 &0.000005 &32.739 &0.434 &0.061 &0.002 &3.169 &10.332\\
\hline
ASASSN-17qj (s36)&F1&sor&66.316&0.27088 &0.00115 &3.691628 &0.015670 &1.799 &0.089 &0.996 &0.008 &0.846 &2.128\\
&F2&orb& &6.84717 &0.00180 &0.146046 &0.000038 &1.148 &0.089 &0.603 &0.012 &0.214 &5.363\\
&F3&nsh&&7.14505 &0.00152 &0.139957 &0.000030 &1.362 &0.089 &0.662 &0.010 &0.224 &6.072\\
\hline
ASASSN-17qj (s63)
&F1&sor&72.844&0.30211 &0.00040 &3.310033 &0.004346 &3.228 &0.062 &0.595 &0.003 &0.568 &5.683\\
&F2&orb&&6.83840 &0.00120 &0.146233 &0.000026 &1.070 &0.062 &0.878 &0.009 &0.588 &1.820\\
&F3&nsh& &7.14649 &0.00012 &0.139929 &0.000002 &10.877 &0.062 &0.349 &0.001 &0.590 &18.428\\
&F4&2*nsh&&14.29510 &0.00067 &0.069954 &0.000003 &1.917 &0.062 &0.726 &0.005 &0.185 &10.344\\
\hline
ZTF18acakuxo (s59)
&F1&sor&35.373 &0.24168 &0.00077 &4.137773 &0.013138 &2.132 &0.074 &0.863 &0.006 &0.398 &5.360 \\
&F2&orb&&3.99225 &0.00142 &0.250485 &0.000089 &1.155 &0.074 &0.005 &0.010 &0.257 &4.496 \\
&F3&nsh&&4.23338 &0.00067 &0.236218 &0.000037 &2.443 &0.074 &0.500 &0.005 &0.257 &9.494\\
\enddata
\begin{threeparttable} 
	\begin{tablenotes} 
		\item [a] orb, nsh and sor represent orbital, NSH and SOR signals, respectively, and 2*orb represents: $ f = 2*f_{\rm orb} $.
		
	\end{tablenotes} 
\end{threeparttable}
\end{deluxetable*}


\section{Analysis and Results} \label{sec:Analysis}
\subsection{ASAS J142023-4856.0} \label{subsec:ASAS J142023-4856.0}

\begin{figure}
\centerline{\includegraphics[width=0.9\columnwidth]{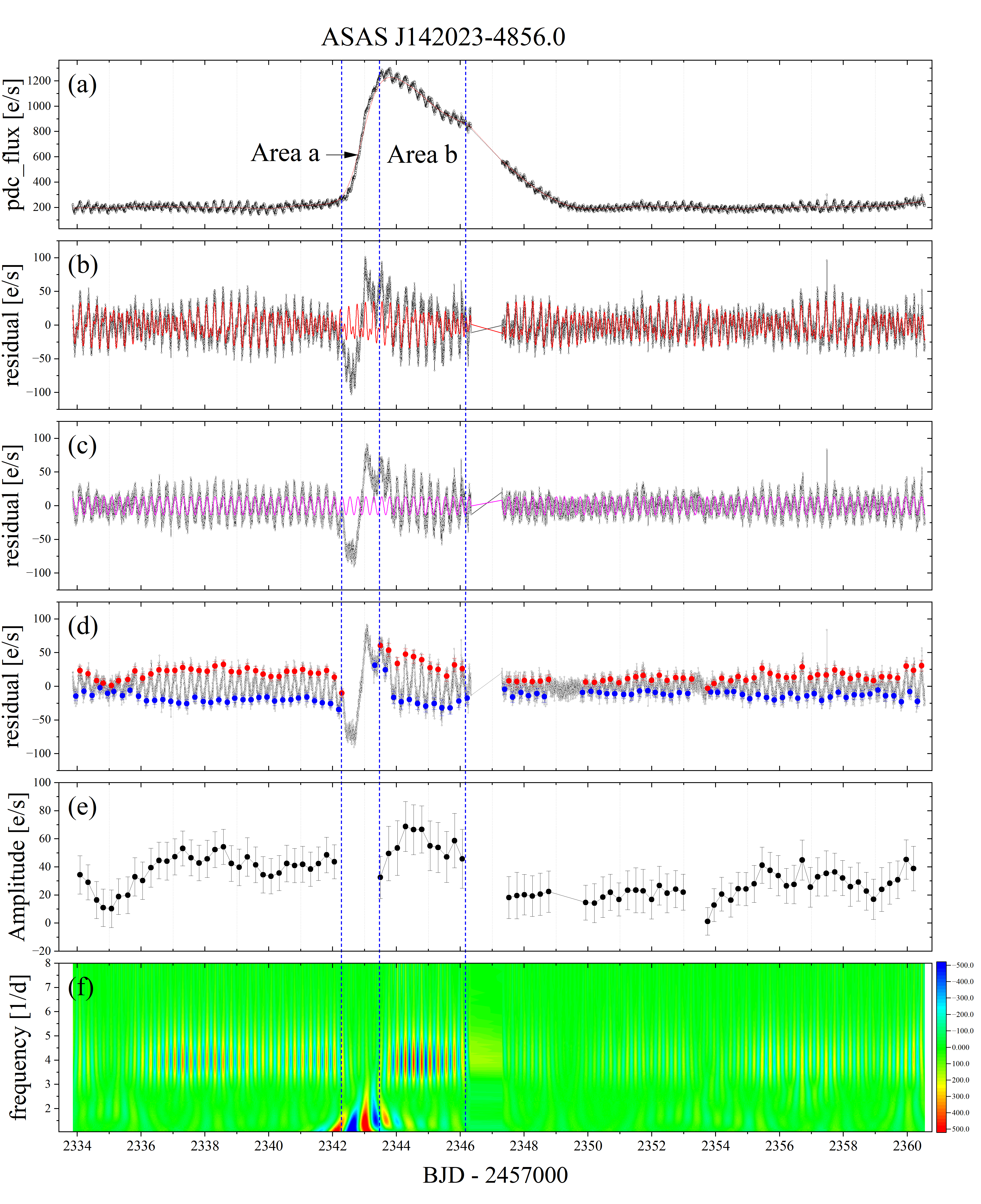}}
\caption{The analysis process of ASAS J1420. Plate (a): the black dots are the light curves of ASAS J1420 from \textit{TESS} photometry, and the red solid line is the LOESS fit; plate (b): the black dots are the residuals of the LOESS fit, and the red solid line is fit to signals with S/N $\textgreater$ 4.0 using Eq. \ref{eq:sine}; plate (c): the black dotted line is the residual (NSH light curve) after removal of the non-NSH signal using Eq. \ref{eq:sine2}, and the magenta solid line is the fitted curve for the NSH signal using Eq. \ref{eq:sine3}; plate (d): an example of using Gaussian fitting to obtain NSH maxima (red dots) and minima (blue dots); plate (e): NSH amplitudes determined using NSH maxima and minima; plate (f): 2D power spectra obtained by CWT of the residual light curve after the removed non-NSH signal. \label{142023}}
\end{figure}

ASAS J142023-4856.0 (hereafter ASAS J1420) was first discovered by \cite{2013Gulbis} as a very bright CV (up to 12 magnitudes), but no other studies followed. A normal outburst of ASAS J1420 was observed in TESS sector 38, with an outburst duration of about 10 d (see Fig. \ref{142023} a). For the analysis of ASAS J1420, we use the following steps:

(i) A locally weighted regression (LOESS; \citealp{cleveland1979robust}) fit with a span of 0.05 days was used to remove outburst trends;

(ii) \texttt{Period04} software (see \citealp{Period04} for details) was used for frequency analysis, and frequencies with signal-to-noise ratios (S/N) greater than 4.0 were used to fit the light curve and then determine the orbital period, SORs, and NSHs or PSHs.

(iii) Based on the fitting parameters obtained in (ii), the theoretical curve of the non-NSH signals was subtracted from the original curve, and the residuals were used to analyze the NSHs and plot the fitted curve of the NSHs for visualization;

(iv) We fold the residuals after removing the outburst trend and non-NSH signals to 0.0 - 2.0 phases, using the NSH period and a minima as the period and zero phase, respectively.

(v) As in our recent work, Gaussian fitting was used to compute the maxima and minima of each NSH and finally to obtain the amplitude of the NSHs (\citealp{2023arXiv230905891S});

(vi) The light curves after removing outburst trends and orbital signals were analyzed using the Continuous Wavelet Transform (CWT; \citealp{polikar1996wavelet}) to obtain the NSHs over time.

\begin{figure}
\centerline{\includegraphics[width=0.7\columnwidth]{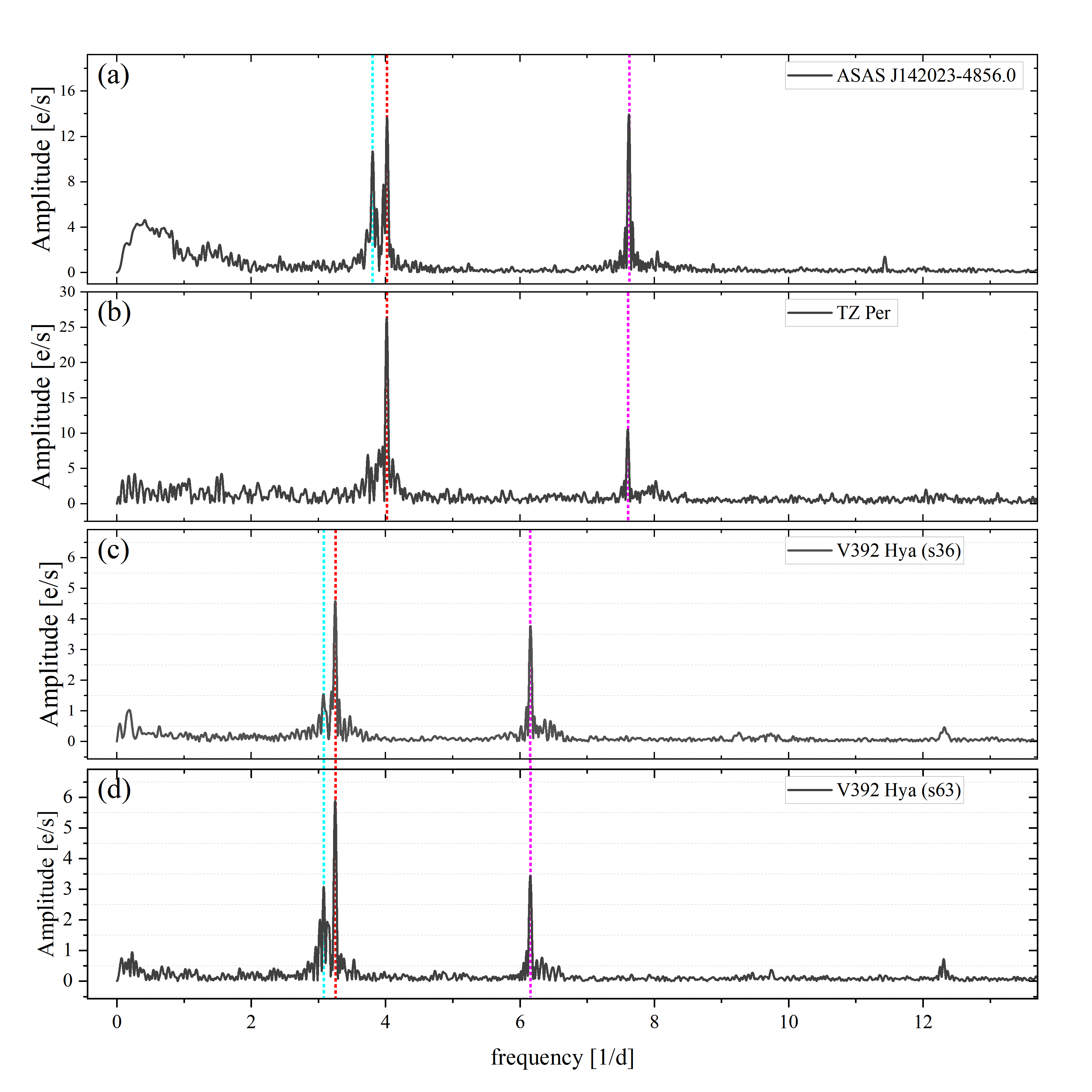}}
\caption{Spectrograms of ASAS J1420, TZ Per, and V392 Hya. The vertical dotted lines in fluorescent, red, and magenta colors correspond to the orbital period, NSHs, and the second harmonic of the orbital period. Plates (c) and (d) correspond to V392 Hya data from sectors 36 and 63, respectively. \label{3lsp}}
\end{figure}

In step (i), PDC\_FLUX was used for the analysis, and a LOESS fit spanning 0.05 days was used to remove outburst trends, followed by frequency analysis of the residuals. Fourier spectra were obtained with the \texttt{Period04}, and signal-to-noise ratios (S/N) were calculated for frequencies with a box size of 2 days $^{-1}$. The results show the presence of four signals (S/N $\textgreater$ 4.0) in the spectra: F1 = 0.262676(24) d, F2 = 0.248654(16) d, F3 = 0.131222(4) d, and F4 = 0.087482(19) d (see Fig. \ref{142023}, Fig. \ref{3lsp} and Tab. \ref{data}). F3 and F4 can be recognized as the second and third harmonics of F1, but it is not immediately clear which of F1 and F2 is the orbital signal. If F1 is an orbital period then F2 is NSHs, if F2 is an orbital period then F1 is PSHs.

Only two (TZ Per and V392 Hya) of the nine stars in this work have had their orbital periods measured by previous authors, leaving seven for which we will be determining the orbital period and the hump period for the first time. PSHs are commonly found during the superoutbursts that characterize SU UMa-type DNe \citep{2009PASJ...61S.395K}, but recent studies have shown that they are also present in the NL \citep{Bruch2022,Bruch2023,BruchIII}. In addition, the accretion disk precession signal has also been found in PSH systems, and we are therefore unable to determine whether the hump belongs to NSHs or PSHs based on the type of star and the presence or absence of the precession signal alone.
	
\begin{figure}
	\centerline{\includegraphics[width=0.6\columnwidth]{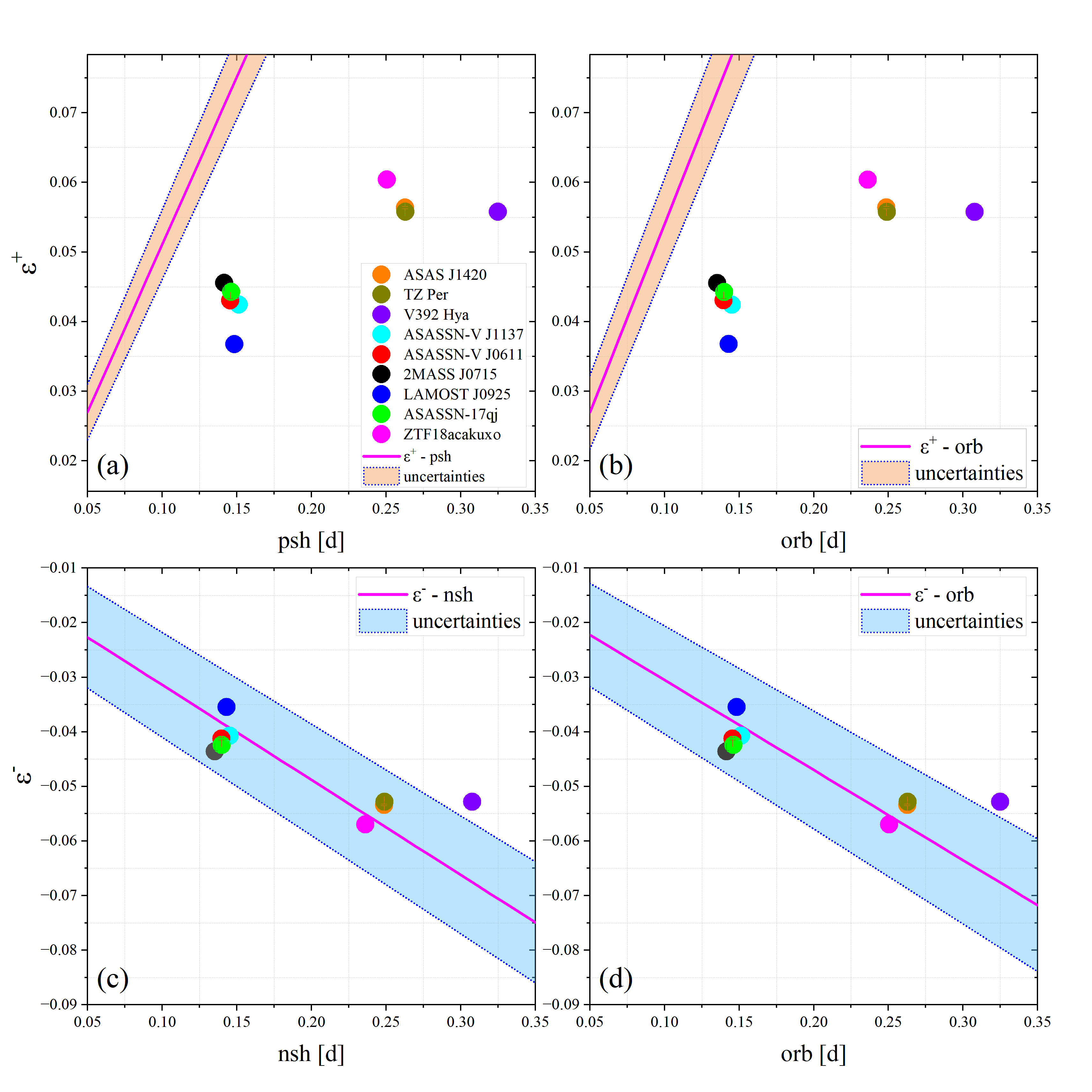}}
	\caption{Comparison between the $\epsilon$ and the four formulas from \cite{BruchIII} after assuming that the humps are PSHs and NSHs, respectively. The solid lines in plates (a) to (d) correspond sequentially to the $\epsilon^{+}$ - $ P_{\rm psh} $, $\epsilon^{+}$ - $ P_{\rm orb} $, $\epsilon^{-}$ - $ P_{\rm nsh} $ and $\epsilon^{-}$ - $ P_{\rm orb} $ relations, and the filled areas around the solid lines are the corresponding uncertainties; the stars corresponding to the different colored solid circles in each plate are listed in the legend for plate (a). See Section \ref{subsec:ASAS J142023-4856.0} for details of the comparison process.\label{3-3}}
\end{figure}

	 Excess is used as a key parameter in the study of hump, with the NSH excess being expressed as $\epsilon^{-} = (P_{\rm nsh} - P_{\rm orb})/P_{\rm orb}$ and the PSH excess as $\epsilon^{+} = (P_{\rm psh} - P_{\rm orb})/P_{\rm orb}$. In a recent study \cite{BruchIII} obtained relationships about $\epsilon^{-}$ - $ P_{\rm nsh} $, $\epsilon^{-}$ - $ P_{\rm orb} $, $\epsilon^{+}$ - $ P_{\rm psh} $, and $\epsilon^{+}$ - $ P_{\rm orb} $ based on counting 74 hump systems (see equations 1 to 4 are from \citealp{BruchIII}, which have been plotted separately in Figs. \ref{3-3}a-d). Therefore, for the seven stars for which no orbital period has been determined, we will determine the orbital period and the type of hump based on the relationship between the excess and the orbital period, NSHs and PSHs. 

In ASAS J1420, we assume that if F2 = 0.248654(16) d is an orbital period, then F1 = 0.262676(24) d is the PSHs and excess is counted as $\epsilon^{+}$ = 0.0564(2), and if F1 is an orbital period, then F2 is the NSHs and excess is counted as $\epsilon^{-}$ = - 0.0534(2). Based on a comparison of the ASAS J1420 calculations and the four empirical formulas obtained by \cite{BruchIII}, it can be found that if F2 is an orbital period, $\epsilon^{+}$ deviates from relations $\epsilon^{+}$ - $ P_{\rm psh} $ and $\epsilon^{+}$ - $ P_{\rm orb} $ to different levels (see Figs. \ref{3-3}a and b), but if F1 is an orbital period then $\epsilon^{-}$ is all within the uncertainty of $\epsilon^{-}$ - $ P_{\rm nsh} $ and $\epsilon^{-}$ - $ P_{\rm orb} $ (see Figs. \ref{3-3}c and d), so we identify F1 as NSHs.

For the remaining stars for which no orbital period was determined we used the same method as for ASAS J1420. In all cases the period excess falls within (or nearly so) the uncertainty range of the $\epsilon^{-}$ - $ P_{\rm nsh} $ and $\epsilon^{-}$ - $ P_{\rm orb} $ relation of \cite{BruchIII}, but not of his $\epsilon^{+}$ - $ P_{\rm psh} $ and $\epsilon^{+}$ - $ P_{\rm orb} $ relation. Therefore, we categorize all hump signals as NSHs.

\texttt{Period04} provides a least squares fit using Eq. \ref{eq:sine} (see \citealp{Period04} for details):
\begin{equation} \label{eq:sine}
	\rm Flux(t) = Z + \Sigma A_i*\sin(2\pi*(\Phi_i*t+\Omega_i))
\end{equation}
$\rm Z$, $\rm A_{\rm i}$, $\rm \Phi_{\rm i}$ and $\Omega_{\rm i}$ are the fitted intercept, amplitude, frequency, and phase, respectively.
Frequencies with S/N $\textgreater$ 4.0 were fitted by Eq. \ref{eq:sine} to the light curve of the de-outburst trend (see Fig. \ref{142023} b), and the best-fit parameters are shown in Tab. \ref{data}. To study the NSHs in more detail, we remove the orbital signals using the equation:

\begin{equation} \label{eq:sine2}
	\rm residual(t) = light curve (t) - [ Z + \Sigma A_i*\sin(2\pi*(\Phi_i*t+\Omega_i))]
\end{equation}
Parameters $\rm A_{\rm i}$, $\rm \Phi_{\rm i}$ and $\Omega_{\rm i}$ come from signals unrelated to NSHs.
The non-NSH parameters from the ASAS J1420 correspond to F2, F3, and F4 in Tab. \ref{data}.
Therefore, we investigated NSHs alone using the residual light curve of the original curve subtracted from the non-NSH signal using Eq. \ref{eq:sine2}. The residual curves we show twice in Fig. \ref{142023}, the first time to fit the NSHs alone (see Fig. \ref{142023} c), and the second time to show the calculations for the maxima and minima (see Fig. \ref{142023} d).
To visualize NSHs, we plotted the fitted curve of NSH using the equation (see solid magenta line in Fig. \ref{142023} c):
\begin{equation} \label{eq:sine3}
	\rm NSH(t) =  A_{\rm nsh}*\sin(2\pi*(\Phi_{\rm nsh}*t+\Omega_{\rm nsh}))
\end{equation}
The light curves after removal of the non-NSH signals were folded with a period of 0.248654(16) d using a minima of the NSHs as the zero-phase point ($ T_{0} $ = BJD2459337.43832; the folding periods and zero-phase points for the nine stars are listed in Table \ref{data2}), and for a more intuitive presentation, we binned the folded curves with a width of 0.01 phases (see Fig. \ref{FOLD}). There are three stars in this paper from which the outburst trend needs to be removed, and for standardization we carefully and uniformly used a LOESS fit with a span of 0.05 days to remove the outburst trend. There are some data points away from the sinusoidal fit in the folded curve of ASAS J1420 (see Fig. \ref{FOLD}), which are remnants of some of the outburst trend (see Fig. \ref{142023} c: Area a and b), but do not affect the overall NSH trend.

We use the same approach as in recent work \citep{2023arXiv230905891S}, using Gaussian fitting to compute the maxima and minima of each NSH in the light curve after removal of the orbital signal (see Fig. \ref{142023} d), and ultimately to obtain the amplitude of each NSH (see Fig. \ref{142023} e). Finally, the CWT was used to analyse the light curve after removal of the non-NSH signal to obtain the NSH amplitude's evolution with time (see Fig. \ref{142023} f). CWT is an ideal tool to study the variation of NSH signal intensity with time, in Fig. \ref{142023} f the vertical coordinate is the frequency, the horizontal coordinate is the time, and the color scale indicates the intensity of the signal. It can be found that the intensity of NSH varies with time, the stripes of the CWT 2D spectra gradually become redder when the NSH flux increases, and the color changes towards the blue side when the NSH flux decreases. The variation of the CWT 2D spectra is verified with the NSH amplitude obtained by our calculations using the maxima and minima.

Combining the calculated amplitudes and CWT 2D spectra reveals the following characteristics of the NSH amplitude in ASAS J1420: (a) The amplitude of the NSH changes during the quiescence (see Fig. \ref{142023} e and f), but no periodic variation similar to that of the AH Her is detected; (b) The NSHs are undetectable during the rise of the outburst (see Fig. \ref{142023}: Area a ); (c) the amplitude of the NSHs are largest at the peak of the outburst and in the subsequent plateau, and then decreases with the outburst decline (see Fig. \ref{142023}: Area b ).

\subsection{TZ Per} \label{subsec:TZ Per}

TZ Per is a Z Cam-type DN, and \cite{1912Esterre} provided the earliest observations.
\cite{1984PASPSzkody} based on American Association of Variable Star Observers (AAVSO) data found that outburst intervals ranged from 4 to 27 d, averaging 17 d.
\cite{1995MNRASRingwald} determined the orbital period as 6.2520(96) h based on the radial velocity study of the H$\alpha$ emission line, and similarly \cite{1999RMxAA..35..135E} determined the orbital period more precisely as 0.2629062(8) d based on the radial velocity analysis.

TZ Per was observed in the TESS 58 sector, and two total outbursts were incomplete. Although the outbursts are incomplete, it is still possible to distinguish the presence of both long and short outbursts, with the short outbursts having an interval of about 11.60 d (between outburst peaks BJD2459896.72 and BJD2459908.33). Such a short recurrence time is consistent with the outburst characteristics of the Z Cam \citep{pyrzas2012hs}. 

\begin{figure}
\centerline{\includegraphics[width=0.7\columnwidth]{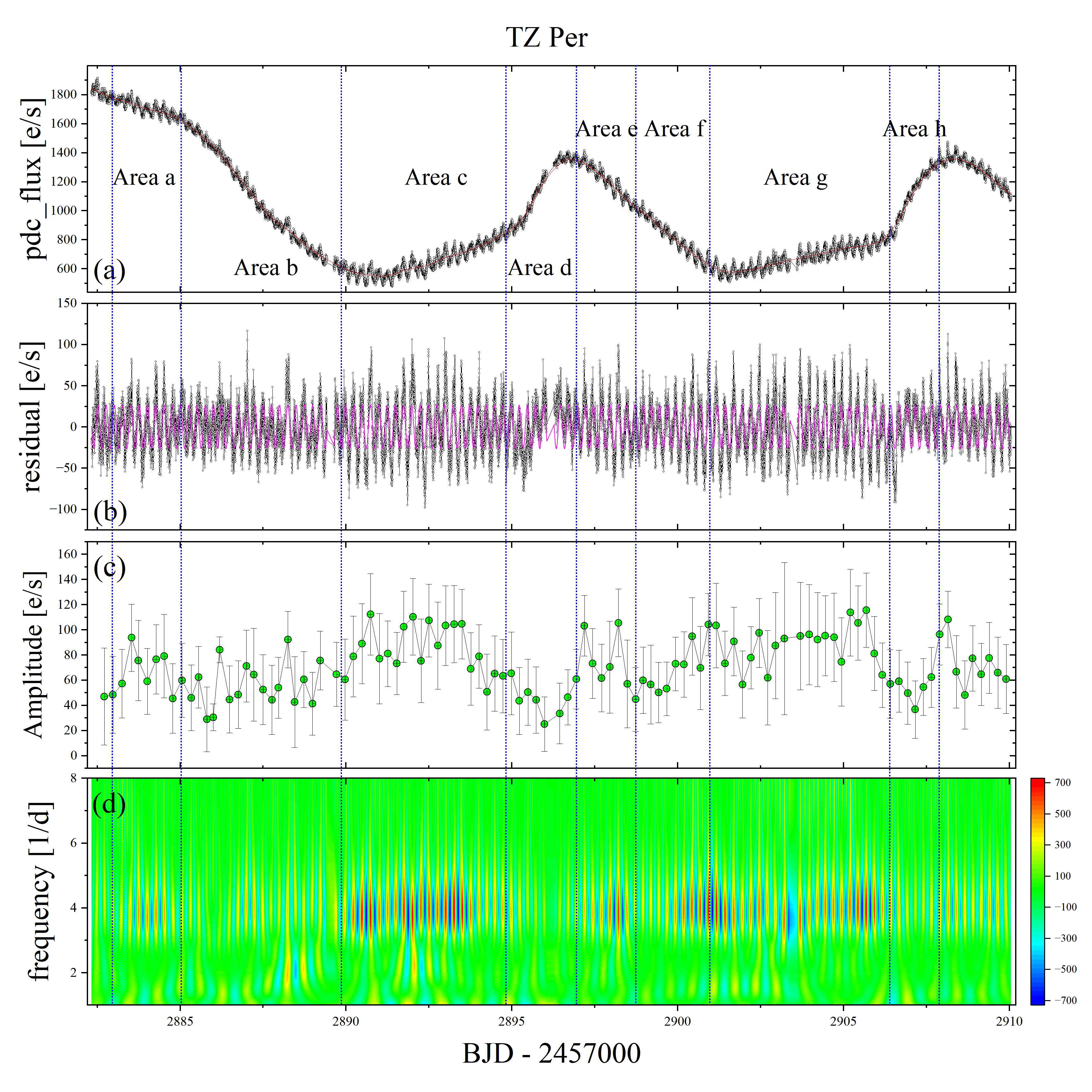}}
\caption{Light curves and amplitude changes in TZ Per. Plate (a): the black dots are the light curves of TZ Per from TESS photometry, and the red solid line is the LOESS fit; plate (b): the black dots are the light curve of the NSH signal, and the magenta solid line is the fitted curves to the NSH signals; plate (c): NSH amplitudes determined using NSH maxima and minima; plate (d): 2D power spectra obtained by CWT of the light curve of the NSH signal. \label{fig:TZ Per}}
\end{figure}

The analytical steps for TZ Per were completely consistent with ASAS J1420. The spectra are shown in Fig. \ref{3lsp}, the light curves after removal of the outburst trend and orbital signals are shown in Fig. \ref{fig:TZ Per} b, and the amplitude variations and CWT 2D spectra are shown in Figs. \ref{fig:TZ Per} c and d, respectively.
The results of the frequency analysis indicate the presence of two significant signals, according to \cite{1999RMxAA..35..135E} obtained for the orbital period, it can be determined that 0.249018(11) d is the NSH period and 0.131509(8) d is the second harmonic of the orbital period. Thus, for the first time, we have determined the orbital period of TZ Per from photometry to be 0.263018(16) d, which is close to \cite{1999RMxAA..35..135E} result, while the NSH excess was determined to be $\epsilon $ = - 0.05283(5). The NSH folding curves using a period of 0.249018(11) d are shown in Fig. \ref{FOLD}.

Based on the NSH amplitude and the CWT 2D spectra, the following features can be found as well: (a) There is a rebound of the NSH amplitude in the recession phase of both long and short outbursts (see Fig. \ref{fig:TZ Per}: Area a  and e);
(b) After passing through the plateau, the NSH amplitude gradually increases as the outburst declines (see Fig. \ref{fig:TZ Per}: Area b and f);
(c) The largest NSH amplitude occurs in quiescence (see Fig. \ref{fig:TZ Per}: Area c and g), but the flux in the quiescence is slowly rising, and the slope of the rise is not as large as in Area d and h. We suggest that the relationship between NSH amplitude and flux can be used to judge the DNe outburst situation.
(d) The NSH amplitude weakens as the outburst rises (see Fig. \ref{fig:TZ Per}: Area d and h).

\subsection{V392 Hya} \label{subsec:MASTER OT J142023.47-485540.0}

V392 Hya was discovered by \cite{2001MNRASChen} during the Edinburgh-Cape Blue Object Survey \citep{1997MNRASStobie} and classified as CV based on spectral characteristics. Kato et al. (2018)(vsnet-chat 8096)\footnote{http://ooruri.kusastro.kyoto-u.ac.jp/mailarchive/vsnet-chat/8096} considers V392 Hya to be a Z Cam-type DN , based on the All-Sky Automated Survey for Supernovae (ASAS-SN) Sky Patrol data \citep{2017PASPKochanek}. \cite{2005PASPPeters} determined the orbital period based on the radial velocity fit as 0.324952(5) d.

\begin{figure}
\centerline{\includegraphics[width=0.7\columnwidth]{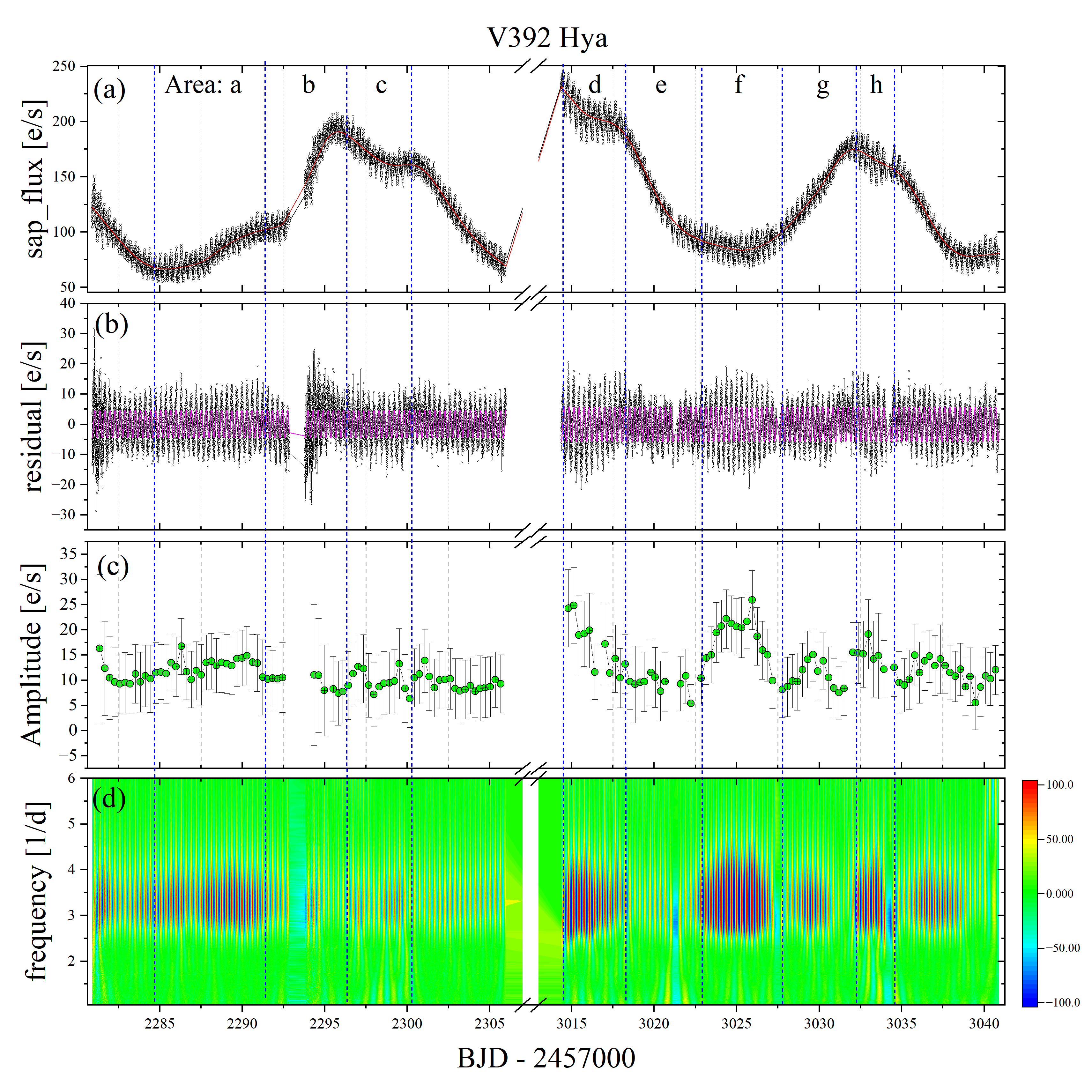}}
\caption{Light curves and amplitude changes in V392 Hya. The individual plates and graphic representations are consistent with TZ Per (see Fig. \ref{fig:TZ Per}).\label{V392 Hya}}
\end{figure}

Three outbursts of V392 Hya were observed in TESS sectors 36 and 63. Since the PDC\_FLUX data partially removes the outburst trend, we use the SAP\_FLUX data to analyze V392 Hya. The light curves of the two sectors were analyzed separately using the same method as ASAS J1420. The results of the analysis for \texttt{Period04} are shown in Table \ref{data} and Fig. \ref{3lsp}, the light curves after removal of the outburst trend and orbital signals are shown in Fig. \ref{V392 Hya} b, and the amplitude variations and CWT 2D spectra are shown in Figs. \ref{V392 Hya} c and d, respectively.
Separate frequency analysis of both sectors reveals the presence of three signals in both.
From the orbital period determined by \cite{2005PASPPeters}, it can be determined that the orbital period based on TESS photometry is 0.325051(52) d (averaged over two sectors) and that the NSH is 0.307784(18) d so that NSH excess is calculated as $\epsilon^{-} $ = -0.05283(7). The $\epsilon^{-} $ of V392 Hya slightly deviates from the uncertainty region of the $\epsilon^{-}$ - $ P_{\rm nsh} $ and $\epsilon^{-}$ - $ P_{\rm orb} $ relations, the orbital period has been previously determined, and therefore the hump type is identified as NSHs.

\begin{figure}
\centerline{\includegraphics[width=\columnwidth]{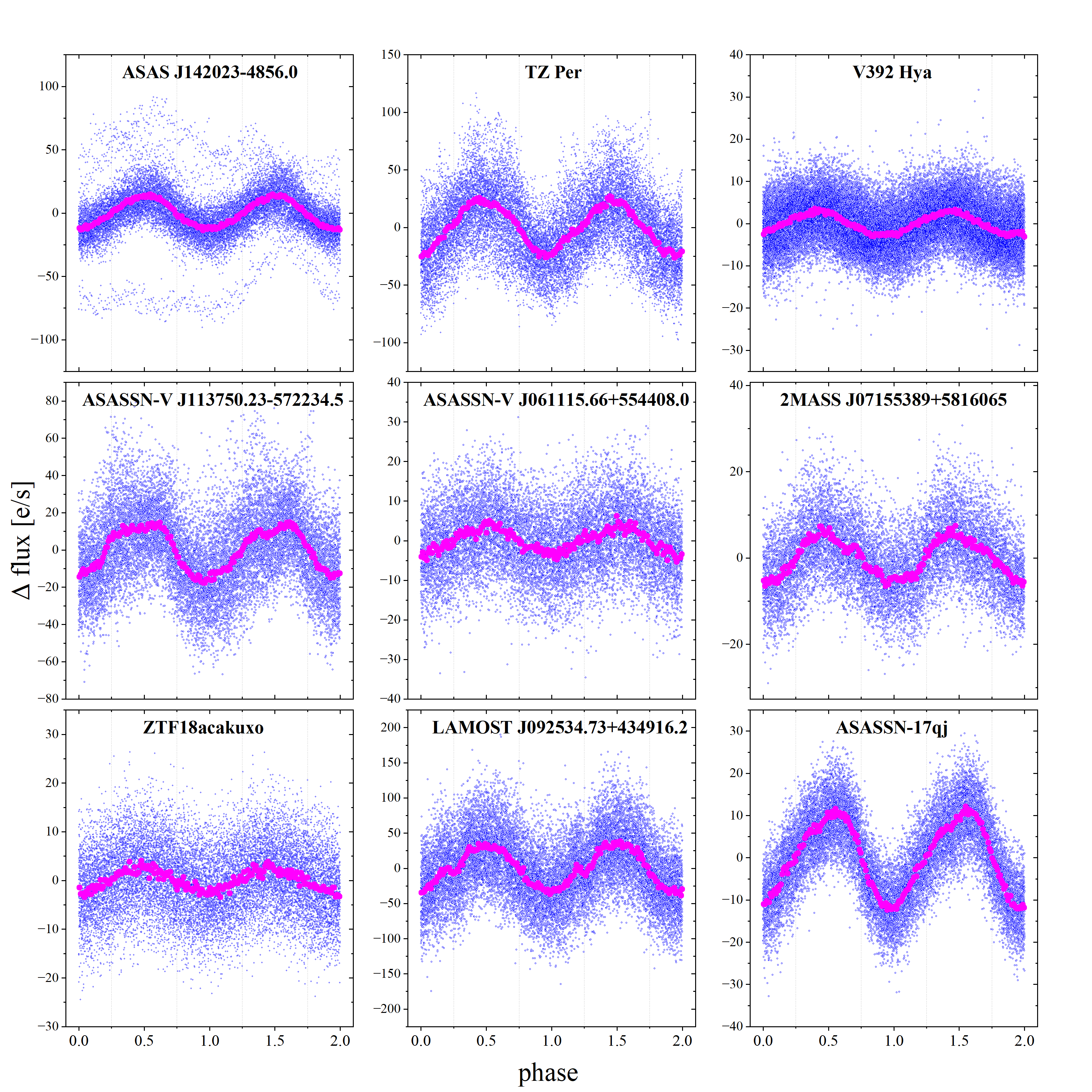}}
\caption{The folded curve of the NSH of nine stars. The folding period and zero-phase points are shown in Table \ref{data2}, and the magenta dots are binned curves with a width of 0.01 phases to show the NSH trend.\label{FOLD}}
\end{figure}

The NSH folding curves using a period of 0.307784(18) d are shown in Fig. \ref{FOLD}.

Based on the calculated amplitude and CWT, as well as frequency analysis, the following characteristics can be obtained:
(a) The NSH amplitude of sector 36 (4.56(5) e/s) is smaller than that of sector 63 (5.75(4) e/s);
(b) Consistent with ASAS J1420, TZ Per, there was a rebound in NSH amplitude after the peak of the outburst (see Fig. \ref{V392 Hya}: Area c, d, and h). 
(c) The NSH amplitude maximum occurs close to the neighborhood of the minimum of the light curve (see Fig. \ref{V392 Hya}: Area a and f). The weakening of the NSH increase as the outburst rises is insignificant due to missing data (see Fig. \ref{V392 Hya}: Area b and g).


\subsection{ASASSN-V J113750.23-572234.5} \label{subsec:ASASSN-V J1137}

ASASSN-V J113750.23-572234.5 (hereafter ASASSN-V J1137) was first discovered by All-Sky Automated Survey for Supernovae (ASAS-SN; \citealp{2018MNRASJayasinghe}), with a median magnitude of 14.7 mag in V-band. Kato et al. (2018)(vsnet-chat 8184)\footnote{http://ooruri.kusastro.kyoto-u.ac.jp/mailarchive/vsnet-chat/8184} suggests that ASASSN-V J1137 be classified as a NL.
ASASSN-V J1137 were separately photometered by TESS in sectors 37 and 64, respectively. We only used the analysis steps (ii) - (iv) consistent with ASAS J1420 to analyze the two sectors separately.

\begin{figure}
\centerline{\includegraphics[width=0.6\columnwidth]{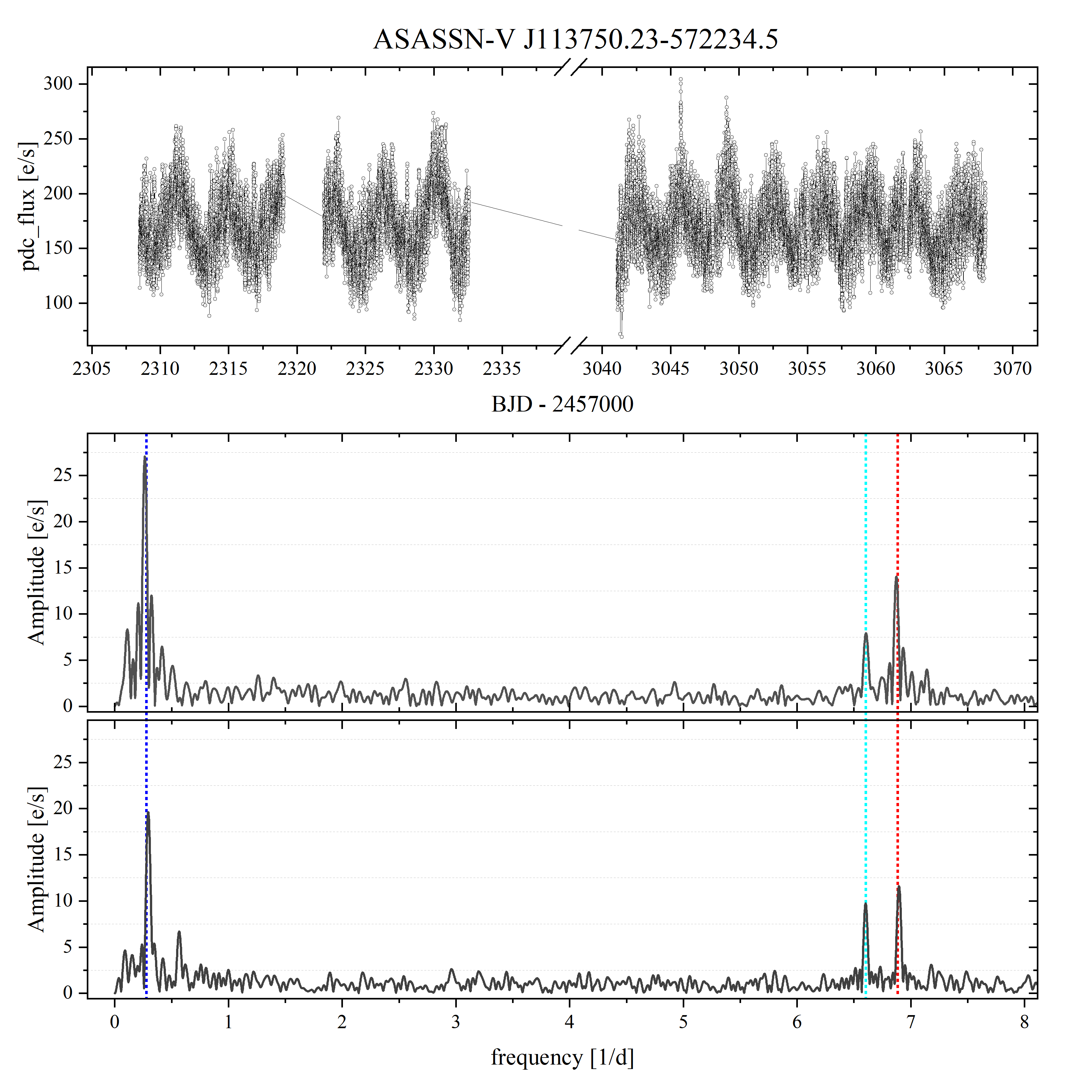}}
\caption{ASASSN-V J1137 light curve and frequency spectrogram. The vertical dotted lines in fluorescent, red, and magenta colors correspond to the orbital period, NSH, and the second harmonic of the orbital period. The vertical dashed lines in blue, fluorescent color, and red correspond to the SOR, orbital period, and NSH, respectively. The middle and bottom panels correspond to sectors 37 and 64, respectively.\label{fig:ASASSN-V J1137}}
\end{figure}

The frequency analysis results are shown in Fig. \ref{fig:ASASSN-V J1137} and Tab. \ref{data}. Therefore, the SOR, orbital period, and NSH of sector 37, are determined to be 3.777(3) d (AM$_{\rm sor, s37}$ = 26.33(21) e/s), 0.151389(14) d (AM$_{\rm orb, s37}$ = 8.03(21) e/s), 0.145499(7) d (AM$_{\rm nsh, s37}$ = 14.22(21) e/s), respectively (see Tab. \ref{vs}). 
In sector 64, the SOR, orbital period, and NSH are determined to be 3.402(3) d (AM$_{\rm sor, s64}$ = 19.51(21) e/s), 0.151449(10) d (AM$_{\rm orb, s64}$ = 10.04(21) e/s), 0.145008(8) (AM$_{\rm nsh, s64}$ = 11.83(21) e/s), respectively. NSH excess were calculated as $\epsilon^{-} _{\rm s37}$ = -0.0389(1) and $\epsilon^{-} _{\rm s64}$ = -0.0452(1), respectively  (see Tab. \ref{vs}). 

Based on the data from sector 37, the orbital period of the data from sector 63 has increased by 0.0044 $ \% $, NSH has decreased by 0.34 $ \% $, and SOR has decreased by 9.93 $ \% $, NSH excess has decreased by 9.32 $ \% $. This small change in orbital period could be real or a measurement error, but we present it here for the purpose of comparison with other signals. It can be seen that all parameters have changed to some extent, except for the orbital period, which has changed less.

Studies of the link between NSHs and SORs a lacking in the literature
NSHs are thought to originate in the tilted disk reverse precession, and thus the link between the SOR and NSH can provide an important window to study the origin of NSHs. Studies of the link between NSHs and SORs a lacking in the literature, and the fact that both NSH and SOR are observed in different sectors provides us with the opportunity to study this topic.
Therefore, we also focus on the changes in SOR and NSH amplitudes. The amplitude of SOR decreases from AM$_{\rm sor, s37}$ = 26.33(21) e/s to AM$_{\rm sor, s64}$ = 19.51(21) e/s, and the NSH amplitude also decreases from AM$_{\rm nsh, s37}$ = 14.22(21) e/s to AM$_{\rm nsh, s64}$ = 11.83(21) e/s). We will make a detailed comparison in Section \ref{subsec:NSH and SOR}.
After removing the non-NSH signals using the same method as in ASAS J1420, the curves are folded using the NSH period, and the folded curves are shown in Fig. \ref{FOLD}.

\subsection{ASASSN-V J061115.66+554408.0} \label{subsec:ASASSN-V J061115.66+554408.0}

ASASSN-V J061115.66+554408.0 (hereafter ASASSN-V J0611) was also first found by the ASAS-SN with an average of 16.5 mag in the V-band, which classified as a U Gem-type DN by the International Variable Star Index (VSX; \citealp{2006SASSWatson}), but no outbursts were observed in TESS photometry.

\begin{figure}
\centerline{\includegraphics[width=0.6\columnwidth]{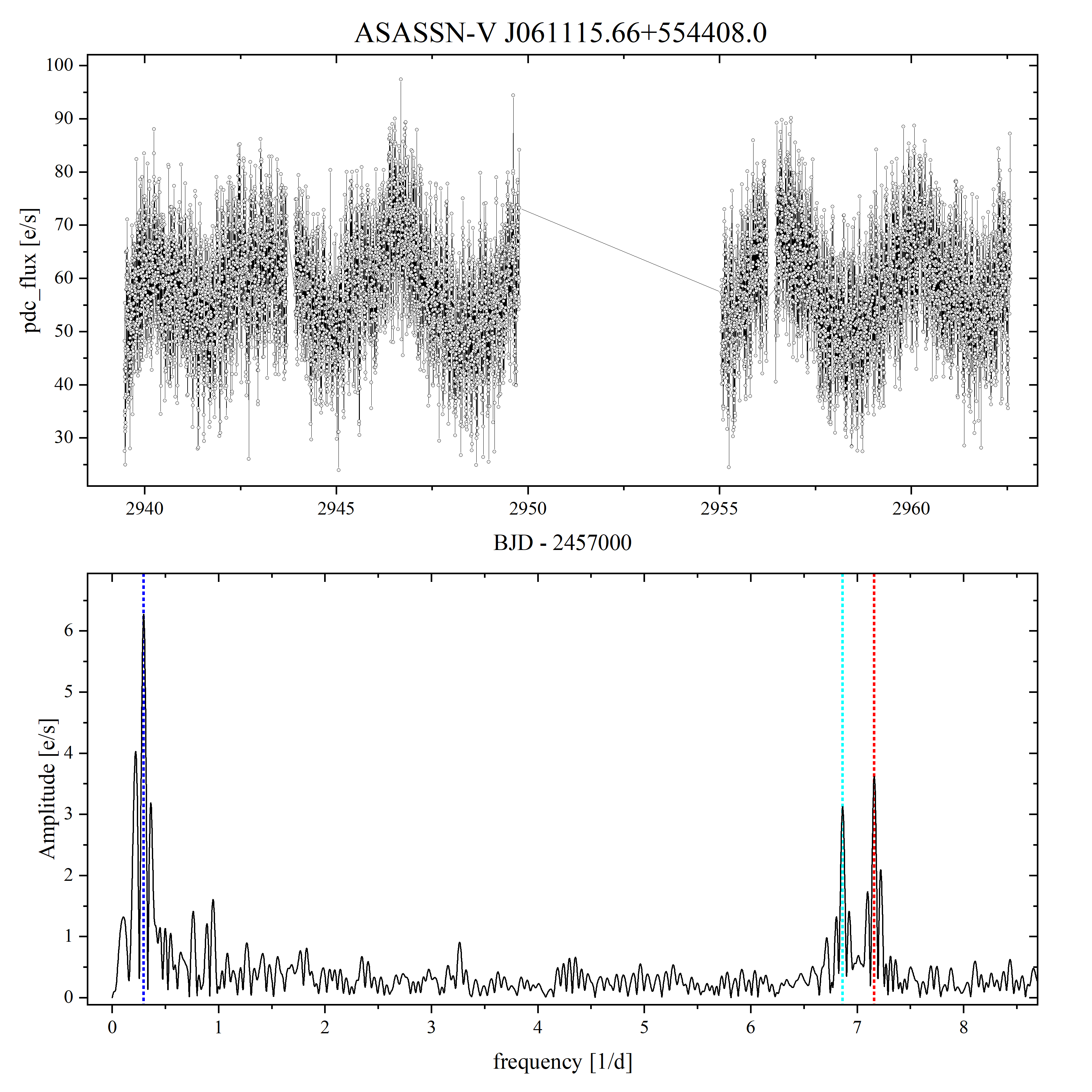}}
\caption{ASASSN-V J0611 light curve and frequency spectrogram. The vertical line of the bottom panel represents the same meaning as in Fig. \ref{fig:ASASSN-V J1137}, and the same vertical line appears in Figs. \ref{fig:2MASS J0715}, \ref{fig:LAMOST J0925}, \ref{fig:ASASSN-17qj} and \ref{fig:ZTF18acakuxo}. \label{fig:ASASSN-V J0611}}
\end{figure}

ASASSN-V J0611 was observed in TESS sector 60. We analyzed ASASSN-V J0611 using methods consistent with ASASSN-V J1137. Based on the results of the frequency analysis (see Fig. \ref{fig:ASASSN-V J0611} and Tab. \ref{data}), the orbital period, NSH, and SOR were determined to be 0.145672(18) d, 0.139660(14) d and 3.359(4) d, respectively. The excess was calculated as $\epsilon^{-}$ = -0.0413(2). Fig. \ref{FOLD} shows the folded curve using the 0.139660(14) d period.

\subsection{2MASS J07155389+5816065} \label{subsec:2MASS J07155389+5816065}

2MASS J07155389+5816065 (hereafter 2MASS J0715) was first observed by the Catalina Real-Time Transient Survey (CRTS; \citep{2009ApJ...696..870D}) in 2006-12-17 with an average magnitude of 16.33 mag in the V-band.
Based on the spectroscopy with the strong He II emission line and other spectral features from the Large Sky Area Multi-Object Fiber Spectroscopic Telescope (LAMOST) survey, \cite{2021ApJS..257...65S} classified 2MASS J0715 as a possible magnetic CV. 
Kato et al. (2018)(vsnet-chat 7970)\footnote{http://ooruri.kusastro.kyoto-u.ac.jp/mailarchive/vsnet-chat/7970} considers 2MASS J0715 to be a VY Scl-type NL . 

\begin{figure}
\centerline{\includegraphics[width=0.6\columnwidth]{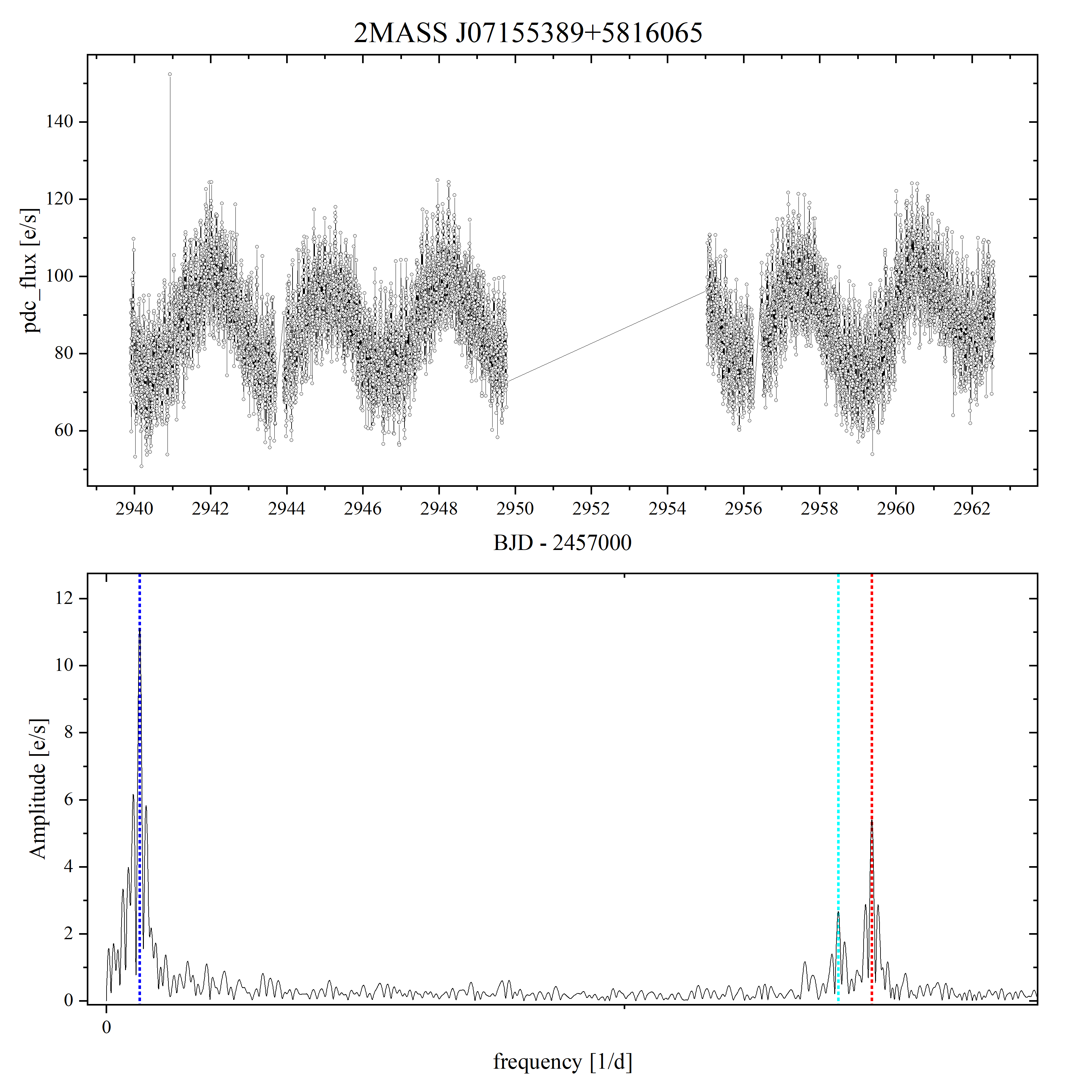}}
\caption{2MASS J0715 light curve from TESS photometry and frequency spectrogram. \label{fig:2MASS J0715}}
\end{figure}

2MASS J0715 was photometric in TESS sector 60, and the results of the frequency analysis of PDC$\_$FLUX showed the presence of three main signals: 3.106(2) d, 0.141498(16) d and 0.135333(7) d, which can be identified in order as the SOR, orbital period and NSH according to the relationship between the three (see Fig. \ref{fig:2MASS J0715} and Tab. \ref{data}). The excess was calculated as $\epsilon^{-}$ = - 0.0436(2). The folding curves of NSH are shown in Fig. \ref{FOLD}.

\subsection{LAMOST J092534.73+434916.2} \label{subsec:LAMOST J092534.73+434916.2}

\begin{figure}
\centerline{\includegraphics[width=0.6\columnwidth]{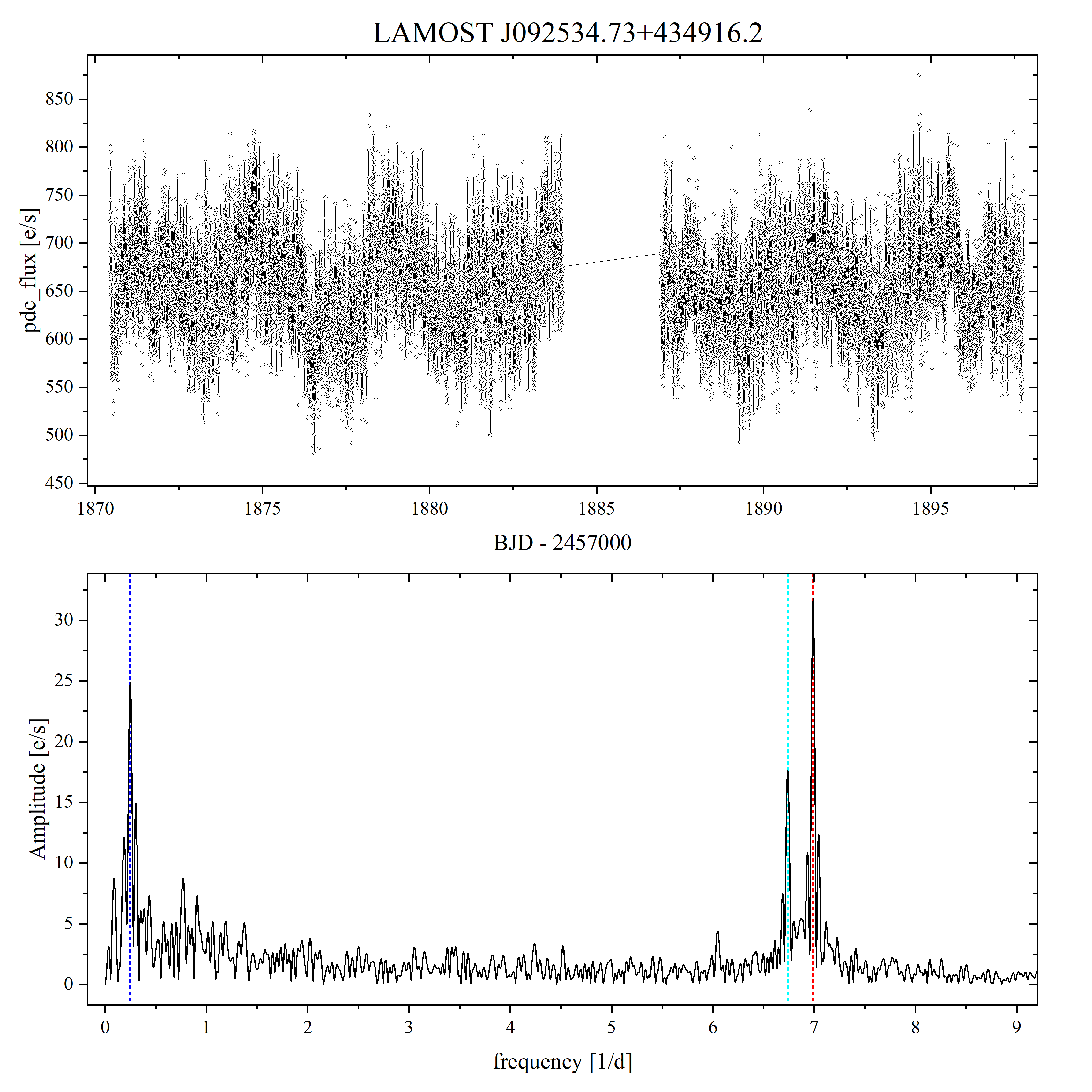}}
\caption{LAMOST J0925 light curve from TESS photometry and frequency spectrogram. \label{fig:LAMOST J0925}}
\end{figure}

LAMOST J092534.73+434916.2 (hereafter LAMOST J0925) was classified by \cite{2020AJ....159...43H} as NL candidates based on The LAMOST survey. LAMOST J0925 was photometric by TESS in sector 21. Based on the frequency analysis (see Fig. \ref{fig:LAMOST J0925} and Tab. \ref{data}), the orbital period, NSH, and SOR were determined to be 0.148348(10) d, 0.143087(5) d and 4.038(6) d, respectively, and the excess was calculated to be $\epsilon^{-}$ = - 0.0355(1).

\subsection{ASASSN-17qj} \label{subsec:ASASSN-17qj}

ASASSN-17qj was first observed by ASAS-SN, classified as a VY Scl-type NL by VSX, and the study was poor. ASASSN-17qj was observed in sectors 36 and 63 of TESS. We perform frequency analysis on the light curves of the two sectors separately. 

\begin{figure}
\centerline{\includegraphics[width=0.6\columnwidth]{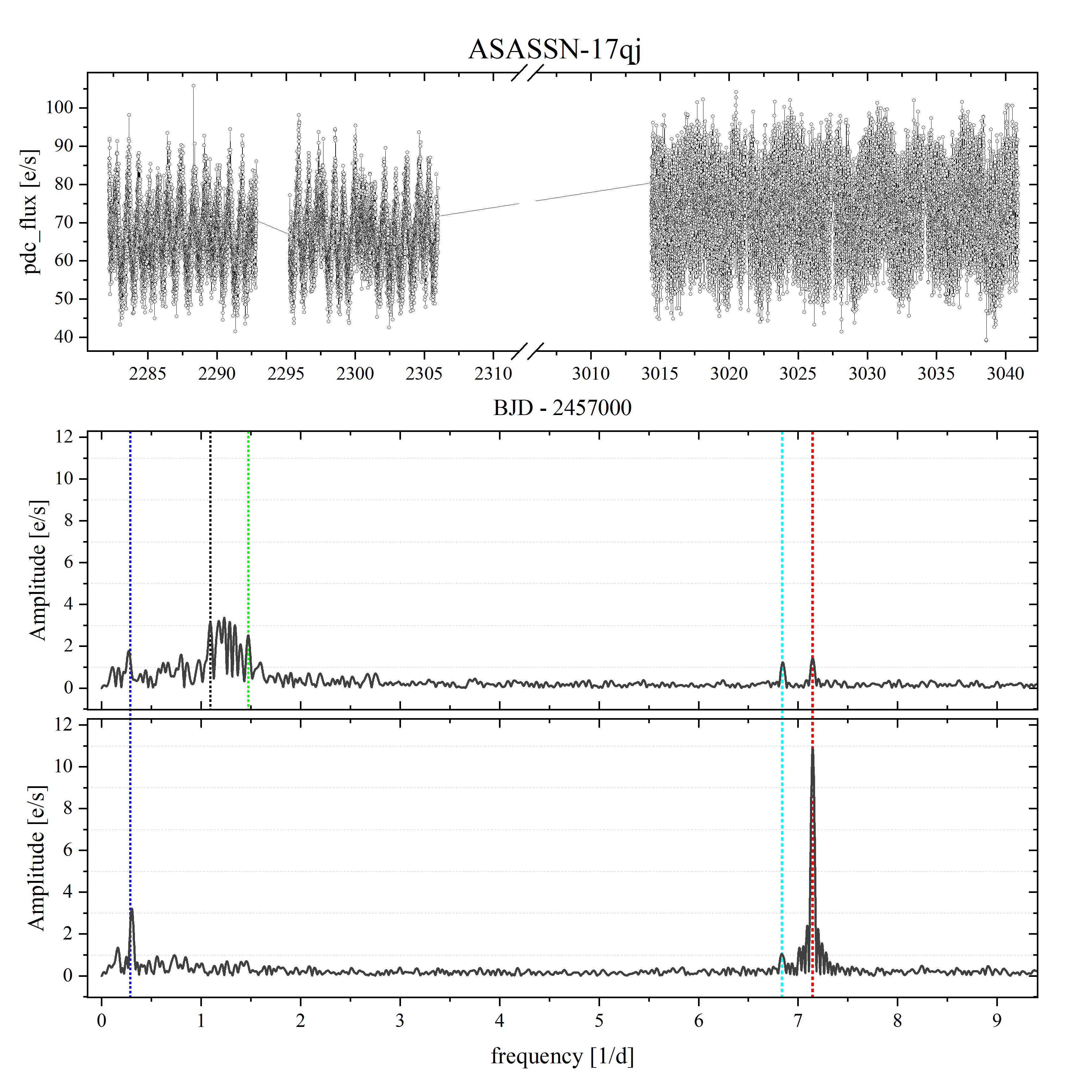}}
\caption{ASASSN-17qj light curve from TESS photometry and frequency spectrogram. The black and green dashed lines in the base plate correspond to 0.9167(5) d and 0.6784(4) d, respectively, and the middle and bottom panels correspond to sectors 36 and 63, respectively. \label{fig:ASASSN-17qj}}
\end{figure}

The frequency analysis results of ASASSN-17qj indicate the presence of an accretion disk precession signal, a hump signal, and a orbital signal in both sectors (see Fig. \ref{fig:ASASSN-17qj}). We used the same method as ASAS J1420 to take the average of the two sectors to determine the type of hump. When the hump signal belongs to the NSHs it is consistent with the  $\epsilon^{-}$ - $ P_{\rm nsh} $ and $\epsilon^{-}$ - $ P_{\rm orb} $ relationship given by \cite{BruchIII}, so we categorize ASASSN-17qj as a NSH system.

Although the S/N of the SOR signal in sector 36 and the orbital signal in sector 63 is less than 4.0, in order to analyze the variations of their amplitudes we counted it still in Tab. \ref{data}. In addition to this, the light curve of ASASSN-17qj in sector 36 shows a periodic variation from 0.6784(4) d to 0.9167(5) d. Due to the low S/N, we do not have any statistics, and in this paper, we mainly concentrate on the analysis of hump and SOR.

In sector 36, SOR, orbital period, NSH and excess are identified as 3.692(16) d (AM$_{\rm sor, s36}$ = 1.80(9) e/s), 0.146046(38) d (AM$_{\rm orb, s36}$ = 1.15(9) e/s), 0.139957(30) d (AM$_{\rm nsh, s36}$ = 1.36(9) e/s) and -0.0417(5) respectively (see Tab. \ref{vs}).
In sector 63, SOR, orbital period, NSH and excess are identified as 3.310(4) d (AM$_{\rm sor, s63}$ = 3.23(6) e/s), 0.146233(26) d (AM$_{\rm orb, s63}$ = 1.07(6) e/s), 0.139929(2) d (AM$_{\rm nsh, s63}$ =  10.88(6) e/s ) and -0.0431(2) respectively (see Tab. \ref{vs}). The mean values for the two sectors were calculated to be 0.146139(32) d, 0.139943(16) d, and 3.501(10) d, respectively, with an excess of -0.0424(3). Because the NSH in sector 63 has a large amplitude, we chose to fold the NSH for this sector using 0.139943(16) d (see Fig. \ref{FOLD}).
The amplitudes of the orbital signals in the two sectors are AM$_{\rm orb, s36}$ = 1.15(9) e/s and AM$_{\rm orb, s63}$ = 1.07(6) e/s, respectively, without significant variations. NSH and SOR have an amplitude of AM$_{\rm nsh, s36}$ =  1.36(9) e/s (AM$_{\rm sor, 36}$ = 1.80(9) e/s) and AM$_{\rm nsh, s63}$ =  10.88(6) e/s (AM$_{\rm sor, s63}$ = 3.23(6) e/s).

If we classify the hump of ASASSN-17qj as NSHs, then there may be a question here whether it is better to classify the hump as PSHs, based on the weaker stability of the orbital period (P$_{\rm orb, s36}$ = 0.146046(38) d; P$_{\rm orb, s63}$ = 0.146233(26) d) than the NSH period (P$_{\rm nsh, s36}$ = 0.139957(30) d; P$_{\rm nsh, s63}$ = 0.139929(2) d) ? We categorize the ASASSN-17qj hump as NSHs for the following reasons:

(i) We consider that the instability of the orbital period in both sectors may not be real, since the S/N ratio of the orbital signal we categorized in sector 36 is 1.820 (S/N $ < $ 4.0), and the orbital period is not significant which may introduce an uncertainty in the period, and the purpose of our still counting it is to make use of its amplitude for the purpose of comparing it with the other signals.

(ii) The amplitude of the orbital signal (AM$_{\rm orb, s36}$ = 1.15(9) e/s; AM$_{\rm orb, s63}$ = 1.07(6) e/s ) is more stable than the NSH amplitude (AM$_{\rm nsh, s36}$ =  1.36(9) e/s; AM$_{\rm nsh, s63}$ =  10.88(6) e/s), and the unstable component we consider to come from the NSH.

(iii) If we assume that the humps are PSHs, then $\epsilon^{+}$ deviates from the $\epsilon^{+}$ - $ P_{\rm psh} $ and $\epsilon^{+}$ - $ P_{\rm orb} $  relation from \cite{BruchIII}, but categorizing the humps as NSHs, $\epsilon^{-}$ coincides with the  $\epsilon^{-}$ - $ P_{\rm nsh} $ and $\epsilon^{-}$ - $ P_{\rm orb} $ relation (see green circles in Figs. \ref{3-3}).

In summary, we believe it is reasonable to categorize the ASASSN-17qj hump as NSHs based on the available information. Of course, it is also hoped that subsequent studies will accurately determine the orbital period and thus verify our choice.

\subsection{ZTF18acakuxo} \label{subsec:ZTF18acakuxo}

\begin{figure}
\centerline{\includegraphics[width=0.6\columnwidth]{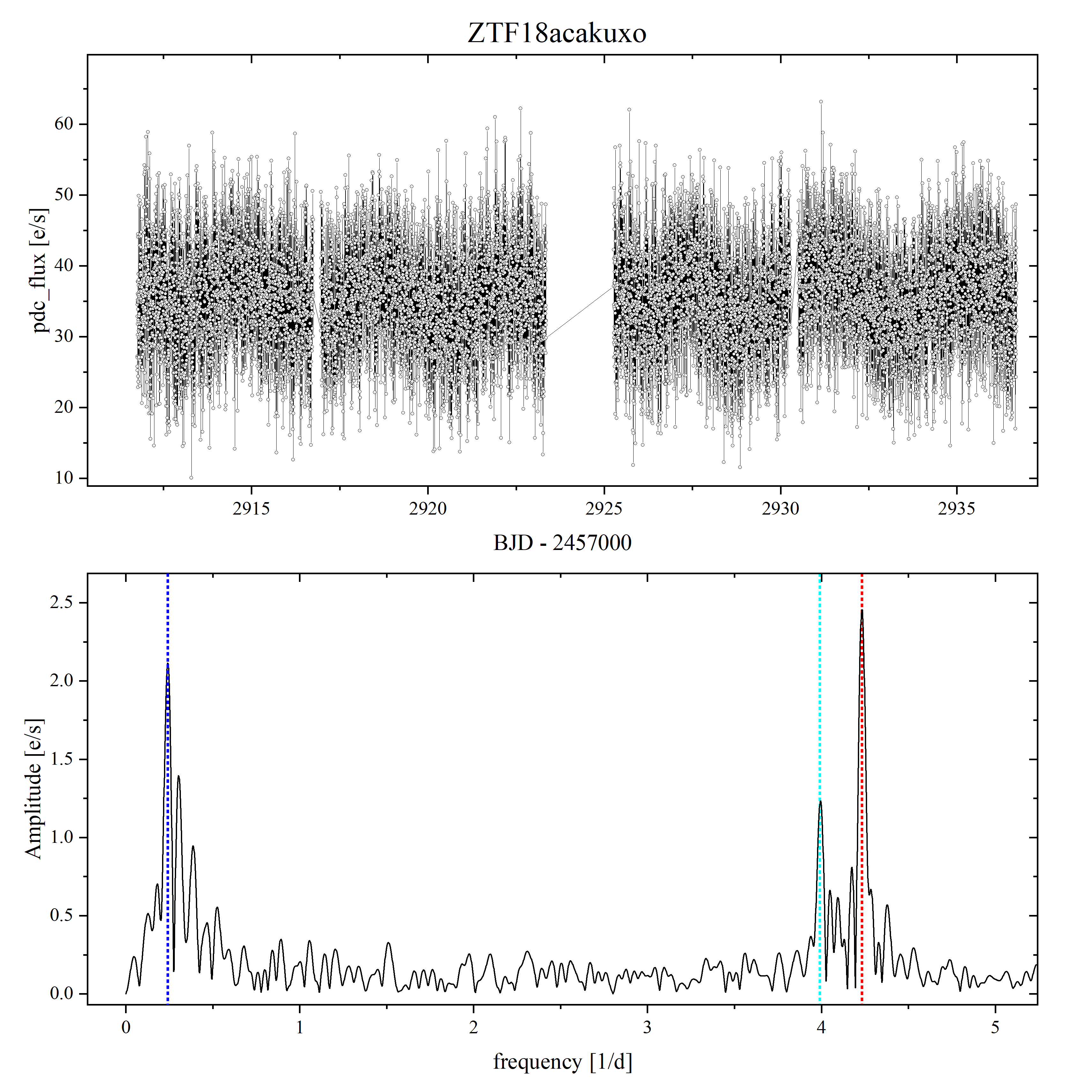}}
\caption{ZTF18acakuxo light curve from TESS photometry and frequency spectrogram. \label{fig:ZTF18acakuxo}}
\end{figure}

ZTF18acakuxo was found by Zwicky Transient Facility (ZTF; \citealp{2019ZTF}) and was classified as CV by ALeRCE \citep{2021AJ....161..242F}, with no follow-up studies. ZTF18acakuxo was observed by TESS sector 59, and based on the frequency analysis also found the presence of SOR, NSH and orbital signals, which were identified as 4.138(13) d, 0.236218(37) d and 0.250485(89)d, respectively (see Fig. \ref{fig:ZTF18acakuxo} and Tab. \ref{data}), so the excess was calculated as - 0.0570(5).

\section{DISCUSSION} \label{sec:DISCUSSION}

\subsection{NSH and DN outbursts} 

In our recent work \citep{2023arXiv230905891S}, we found that NSH was present in both outbursts and quiescence of AH Her and varied with the outburst. The NSH amplitude is most prominent during the quiescence, weakens as the outburst rises, is undetectable at the top, rebounds and weakens during the plateau, and strengthens again as the outburst weakens. Therefore, we suggest that the relationship between NSH amplitudes and outbursts can be used as a window to study accretion disk instability and the origin of NSHs. In the present work, we have newly discovered the presence of NSHs during outbursts in three stars, and the NSH amplitude varies with the outbursts, providing new samples to study the origin of DN outbursts and NSHs. In ASAS J1420, TZ Per, and V392 Hya, there are the following features: (1) The NSH amplitude decreases as the outburst rises and strengthens as the outburst decreases after the plateau; (2) In the plateau following the peak of the outburst, the NSH amplitude rebounds, and this rebound is present in both long and short outbursts; (3) The maximum amplitude of the NSHs in ASAS J1420 occurs during the plateau after the peak outburst, but in TZ Per and V392 Hya it occurs in the quiescence; (4) No SOR-related signals were detected in three stars.

The current study suggests that the tilt angle of the accretion disk \citep{Montgomery2009MNRAS}, the distribution of material on the disk \citep{Wood2011ApJ}, the radius of the disk \citep{Osaki2013b}, and the variation of the material transfer may be related to the NSH amplitude \citep{Smak2013}. Based on our results, it is believed that the NSH amplitude correlates with the stage of the DN outburst. If the accretion disk angle change should be reflected in the light curve, and there is currently no SOR signal in the three stars, the outburst time scale is small, so the change in the accretion disk angle is not considered here.
The release of gravitational potential energy from the accretion stream impinging on the accretion disk is considered a possible origin of the NSHs. 
Therefore, in the change of NSH amplitude of AH Her, we suggest that the change of the radius of the accretion disk during the outburst plays an important role and that the gravitational potential energy released by the accretion stream impinging on the accretion disk after the accretion disk becomes larger with the rise of the outburst will be less than quiescence. 
The change in NSH amplitude during the rise and decline of the outburst is explained by the change in the radius of the accretion disk during the outburst as if it were reasonable, but it does not explain the rebound in NSH amplitude after the peak of the outburst and the fact that the maximum amplitude sometimes occurs in the rebound zone. Therefore, we conclude that the variation in the radius of a single accretion disk is insufficient to explain the variation of the NSH amplitude with DN outbursts and may need to be combined with DIM and consider contributions such as changes in accretion disk material distribution and material transfer. We suggest that DIM model improvement must consider the variability of the NSH amplitude with DN outbursts, particularly the location and extent of the accretion stream reacting with the accretion disk.

\subsection{NSH and SOR} \label{subsec:NSH and SOR}

Orbital, NSHs and SOR signals are predominantly present in NSH systems, but SOR is not detected in all NSH systems. The SOR is considered a periodic change in luminosity due to the reverse precession of the nodal line from the tilted disk, and the SOR period corresponds to the precession period of the tilted disk. The NSHs are thought to originate from the interaction of the reverse precession of the tilted disk with the accretion stream from the secondary star that changes with the orbital motion of the binary. However, there is no definitive conclusion on how the accretion disk is tilted, why it precession in the reverse direction, and how it interacts with the accretion stream. ASASSN-V J1137 and ASASSN-17qj were observed in two TESS sectors, respectively, and the SOR, NSHs, and excess changed to different levels in different sectors, so we made a separate list of comparisons (see Tab. \ref{vs}). 

In ASASSN-V J1137, the amplitude of SOR decreases from AM$_{\rm sor, s37}$ = 26.33(21) e/s to AM$_{\rm sor, s64}$ = 19.51(21) e/s, and the NSH amplitude decreases from AM$_{\rm nsh, s37}$ = 14.22( 21) e/s to AM$_{\rm nsh, s64}$ = 11.83(21) e/s in the case of increasing amplitude of orbital signal (AM$_{\rm orb, s37}$ = 8.03(21) e/s to AM$_{\rm orb, s64}$ = 10.04(21) e/s) (see Tab. \ref{vs}). While the amplitude of the orbital signal increased, the amplitude of both SOR and NSHs decreased.
In ASASSN-17qj, contrary to ASASSN-V J1137, the amplitude of the SOR increases from AM$_{\rm sor, s36}$ = 1.80(9) e/s to AM$_{\rm sor, s63}$ = 3.23(6) e/s, and the amplitude of the NSH increases from AM$_{\rm nsh, s36}$ = 1. 36(9) e/s to AM$_{\rm nsh, s63}$ = 10.88(6) e/s in the case of a slight decrease in the amplitude of the orbital signal (AM$_{\rm orb, s36}$ = 1.15(9) e/s to AM$_{\rm orb, s63}$ = 1.07(6) e/s) (see Tab. \ref{vs}). When the amplitude of the orbital signal decreases, the NSH amplitude increases with the SOR amplitude. Combining the changes in the amplitude of SOR with NSHs for ASASSN-V J1137 and ASASSN-17qj, it is known that such changes are independent of the orbital signal, which may be due to a change in the disk radius, tilt angle, material transfer rate or material distribution.

In addition, the SOR period of ASASSN-V J1137 and ASASSN-17qj is reduced, and the NSH period is also reduced (see Tab. \ref{vs}). 
The frequency of the tilted disk precession ($ \nu_{\rm prec} $) can be expressed as \citep{1995MNRAS.274..987P,1998MNRAS.299L..32L}: 
\begin{equation} \label{eq:cos}
	\nu_{\rm prec} = \frac{1}{P_{\rm sor}}= -\frac{3}{8\pi} \frac{GM_{2}}{a^{3}} \frac{\int \Sigma r^{3}dr}{\int \Sigma\Omega r^{3}dr}  \rm cos\theta
\end{equation}
which has been widely used in the research of the tilted disk precession and NSHs (e.g., \citealp{Osaki2013b}; \citealp{2013PASJ...65...76K}; \citealp{Smak2013}; \citealp{KimuraKIC94066522020PASJ}; \citealp{2023MNRAS.520.3355S}).
In Eq. \ref{eq:cos}, $ G $, $ M_{2} $, $ a $, $ \Sigma $, $ \Omega $, $ r $ and $ \theta $ are the gravitational constant, the mass of the secondary, the binary separation, the surface density of the disk, Keplerian angular velocity of the disk matter, the radial distance from the accretion disk to the white dwarf and tilt angle, respectively. The first thing that can be considered for a large SOR amplitude is a significant change in the inclination angle and radius of the accretion disk or a large matter transfer. Since ASASSN-V J1137 and ASASSN-17qj are both NLs that have been in a thermally stable state for a long time, and since there is no significant amplitude change in the light curves, we assume that the radius of the accretion disk and the material transfer are stable, and consider the change in the inclination angle alone.
\cite{smak2009origin} suggests that the CV's disk tilt should not exceed 7$ ^{\circ} $.  
We assume that the larger amplitude in the SOR in ASASSN-17qj is due to the larger inclination angle of the tilted disk, combined with Eq. \ref{eq:cos} considering only the angle, the precession period is reduced for a larger angle which seems to be consistent with the observational results. As the SOR amplitude increases, the NSH amplitude also increases, and when the inclination angle of the accretion disk increases, the accretion stream will easily enter the inner disk to release more energy, which is consistent with the explanation that the NSH originates from the tilted disk precession.
However, if it is assumed that the decrease in SOR amplitude in ASASSN-V J1137 is due to a decrease in angle, then the precession period is increased, contrary to the observation. Therefore, we can conclude that the variation of the SOR amplitude and period cannot be fully explained by considering the variation of the inclination angle of the accretion disk, and it is still necessary to take into account the variation of the disk radius, the distribution of the material and the transfer of the material.


The relation between $\epsilon ^-$ and $q$ given by \cite{wood2009sph} to estimate the mass ratio of the binary:
\begin{equation} \label{eq:q}
	q = -0.192\mid\epsilon^{-}\mid ^{1/2} + 10.37\mid\epsilon^{-}\mid - 99.83\mid\epsilon^{-}\mid ^{3/2} + 451.1\mid\epsilon^{-}\mid ^{2}
\end{equation}
We calculated the mass ratio separately based on Eq. \ref{eq:q} and the excess of ASASSN-V J1137 and ASASSN-17qj in different sectors. The results show that $ q $ significantly differs for different sectors (see Tab. \ref{vs}), which suggests that excess use to determine the binary's mass ratio needs to be handled with care.

\begin{deluxetable*}{lll}
\tablecaption{Comparison of parameters for different sectors of ASASSN-V J1137 and ASASSN-17qj.\label{vs}}
\tablewidth{0.01pt}
\tabletypesize{\scriptsize}
\tablehead{	
	\colhead{Parameters}&\colhead{Values}&\colhead{Values}}
\startdata
ASASSN-V J1137\\
Sectors&s37&s64\\
Mid-time (BJD-2457000)&2320.544603&3054.576215\\
Orb, period (d)&0.151389(14)&0.151449(10)\\
Orb, amplitude (e/s)&8.03(21)&10.04(21)\\
NSH, period (d)&0.145499(7)&0.145008(8)\\
NSH, amplitude (e/s)&14.22(21)&11.83(21)\\
SOR, period (d)&3.777(3)&3.402(3)\\
SOR, amplitude (e/s)&26.33(21)&19.51(21)\\
$ \epsilon^{-} $, $(P_{\rm nsh} - P_{\rm orb})/P_{\rm orb}$&-0.0389(1)&-0.0425(1)\\
q, M$ _{2} $/M$ _{1} $&0.282&0.341\\
\hline
ASASSN-17qj&\\
Sectors&s36&s63\\
Mid-time (BJD-2457000)&2294.0895&3027.63777\\
Orb, period (d)&0.146046(38)&0.146233(26)\\
Orb, amplitude (e/s)&1.15(9)&1.07(6)\\
NSH, period (d)&0.139957(30)&0.139929(2)\\
NSH, amplitude (e/s)&1.36(9)&10.88(6)\\
SOR, period (d)&3.692(15)&3.310(4)\\
SOR, amplitude (e/s)&1.80(9)&3.23(6)\\
$ \epsilon ^{-}$, $(P_{\rm nsh} - P_{\rm orb})/P_{\rm orb}$&-0.0417(5)&-0.0431(2)\\
q, M$ _{2} $/M$ _{1} $&0.328&0.352\\
\enddata
\end{deluxetable*}

\begin{deluxetable*}{lllllll}
\tablecaption{Parameter statistics for nine NSH systems.\label{data2}}
\tablewidth{0.01pt}
\tabletypesize{\scriptsize}
\tablehead{	
	\colhead{ Stars}&\colhead{P$_{\rm orb}$$ ^{\rm a} $}&\colhead{ P$_{\rm nsh}$$ ^{\rm a} $}&\colhead{ T$_{0}$$ ^{\rm b} $}&\colhead{$ \epsilon^{-} $ }&\colhead{P$_{\rm sor}$$ ^{\rm a} $}&\colhead{Types}\\}
\startdata
ASAS J1420&0.262676(24)&0.248654(16)&2337.43832&-0.0534(2)  &-&DN\\
TZ Per &0.2629062(8)$ ^{\rm c} $&0.249018(11)&2885.42215&-0.05283(5)  &-&DN\\
V392 Hya$ ^{\rm e} $&0.324952(5)$ ^{\rm d} $&0.307784(35)&2282.44505&-0.05283(7)  &-&DN\\
ASASSN-V J1137 $ ^{\rm e} $&0.151419(24)&0.145254(15)&2318.25626&-0.0407(3) &3.590(5)&NL\\
ASASSN-V J0611&0.145672(18)&0.139660(14)&2949.37371&-0.0413(2)  &3.359(4)&DN\\
2MASS J0715&0.141498(16)&0.135333(7)&2949.65513&-0.0436(2) &3.106(2)&NL\\
LAMOST J0925&0.148348(10)&0.143087(5)&1883.69643&-0.0355(1) &4.038(6)&NL\\
ASASSN-17qj$ ^{\rm e} $&0.146139(32)&0.139943(16)&3027.83512&-0.0424(3)  &3.501(19)&NL\\
ZTF18acakuxo&0.250485(89)&0.236218(37)&2920.89540&-0.0570(5) &4.138(13)&CV\\
\enddata
\begin{threeparttable} 
	\begin{tablenotes} 
		\item [a] The unit of the parameter is days.
	\item [b] A minima of the NSHs is used as the zero-phase point for folding the NSH light curves in units \\of BJD - 24547000.
		\item [c] The orbital period comes from \cite{1999RMxAA..35..135E}.	
		\item [d] The orbital period comes from \cite{2005PASPPeters}.
		\item [e] The parameter comes from the average of two sectors.	
	\end{tablenotes} 
\end{threeparttable}
\end{deluxetable*}


\section{CONCLUSIONS} \label{sec:CONCLUSIONS}

In this paper, the negative superhump, superorbital signal, and negative superhump variations with outburst are studied for nine CVs (ASAS J1420, TZ Per, V392 Hya, ASASSN-V J1137, ASASSN-V J0611, 2MASS J0715, LAMOST J0925, ASASSN-17qj, and ZTF18acakuxo) based on TESS photometry, which is summarized as follows:

We find for the first time the existence of NSHs in DNe ASAS J1420, TZ Per, and V392 Hya, with periods determined to be 0.248654(16) d, 0.249018(11) d, and 0.307784(35) d (see Tab. \ref{data2}), respectively. All three stars have DN outbursts, and the NSH amplitudes change with the outbursts, with the most notable features being as follows: (a) the NSH amplitude decreases as the outburst rises; (b) after the peak outburst, the NSH amplitude rebounds at the plateau; (c) as the outburst recedes, the NSH amplitude increases again; (d) during the quiescence, the NSH amplitude is unstable; (e) the maximal NSH amplitude in the different stars is likely to be present at the plateau as well as during the quiescence; (f) No signals related to the accretion disk precession were detected in the three stars. The findings in this paper are similar to and different from AH Her. No periodic changes in the NSH amplitude are found in the new three stars, but in this paper, the NSH amplitude also rebounds after the peak of the short outburst, which provides new observational support for studying the NSH's origin and improving the DN outbursts' model. We suggest that variations in the radius of the accretion disk during outbursts can explain some phenomena regarding variations in the NSH amplitude, such as the rise and fall of outbursts. However, the variation of the accretion disk radius cannot explain the rebound of the NSH amplitude after the peak of the outburst and the fact that the maximum amplitude may no occur in quiescence.
Therefore, we suggest that a single change in the radius of the disk cannot fully explain the variation of NSHs with the outburst and needs to be combined with DIM and considered with other ingredients, which can be used as a window to study the origin of NSHs and to improve the DIM.

The remaining six stars (ASASSN-V J1137, ASASSN-V J0611, 2MASS J0715, LAMOST J0925, ASASSN-17qj, and ZTF18acakuxo), studies are very poor. We confirm for the first time their orbital periods, NSH and SOR, providing a new sample for the study of tilted disk precession models (see Tabs. \ref{data} and \ref{data2}). Especially in ASASSN-V J1137 and ASASSN-17qj, the period and amplitude of SOR and NSH change in different sectors. The NSH amplitude decreases with the decreasing amplitude of SOR in ASASSN-V J1137, and the NSH amplitude increases with the increasing amplitude of SOR in ASASSN-17qj, and the accretion disk precession period decreases for both stars.
Combining the SOR amplitude and period variations of the two stars and the theoretical formulas for the tilted disk precession, we consider only the case of tilted disk angle variations, where an increase in angle can explain the increase in SOR amplitude and decrease in the precession period for ASASSN-17qj. However, a decrease in angle is not consistent with a decrease in the SOR period for ASASSN-V J1137. Therefore, we suggest that considering variations in accretion disk angle does not fully explain the variations in SOR amplitude and period, and that variations in the radius of the accretion disk, material distribution, and material transfer also need to be considered.

In this study, we identified nine hump systems; however, only two (TZ Per and V392 Hya) were classified as NSH systems based on the orbital periods measured by previous researchers. The remaining seven systems were categorized as NSH systems, utilizing the correlation between the excess ($\epsilon$) and the hump periods/orbital periods obtained by \cite{BruchIII}. The most effective method for determining the hump type remains an accurate measurement of the orbital period. Therefore, we would like to emphasize that our classification may be optimistic and require further verification through future work for an accurate determination of the orbital period.

This work was supported by the National Natural Science Foundation of China (Nos. 11933008, 12103084, and 12303040) and the basic research project of Yunnan Province (Grant No. 202301AT070352). The data used in this work comes from publicly available data from TESS mission. Data were downloaded from the Mikulski Archive for Space Telescopes (MAST)\footnote{https://mast.stsci.edu/} and
the ExoFOP-TESS webpage\footnote{https://exofop.ipac.caltech.edu/tess/target.php?id=172518755}.

\bibliography{sample631}{}

\begin{thebibliography}{}
\expandafter\ifx\csname natexlab\endcsname\relax\def\natexlab#1{#1}\fi
\providecommand{\url}[1]{\href{#1}{#1}}
\providecommand{\dodoi}[1]{doi:~\href{http://doi.org/#1}{\nolinkurl{#1}}}
\providecommand{\doeprint}[1]{\href{http://ascl.net/#1}{\nolinkurl{http://ascl.net/#1}}}
\providecommand{\doarXiv}[1]{\href{https://arxiv.org/abs/#1}{\nolinkurl{https://arxiv.org/abs/#1}}}

\bibitem[{{Armstrong} {et~al.}(2013){Armstrong}, {Patterson}, {Michelsen},
  {Thorstensen}, {Uthas}, {Vanmunster}, {Hambsch}, {Roberts}, \&
  {Dvorak}}]{2013MNRAS.435..707A}
{Armstrong}, E., {Patterson}, J., {Michelsen}, E., {et~al.} 2013, \mnras, 435,
  707, \dodoi{10.1093/mnras/stt1335}

\bibitem[{{Balman}(2015)}]{2015AcPPP...2..116B}
{Balman}, S. 2015, Acta Polytechnica CTU Proceedings, 2, 116.
\newblock \doarXiv{1403.4437}

\bibitem[{{Barrett} {et~al.}(1988){Barrett}, {O'Donoghue}, \&
  {Warner}}]{Barrett1988MNRAS}
{Barrett}, P., {O'Donoghue}, D., \& {Warner}, B. 1988, \mnras, 233, 759,
  \dodoi{10.1093/mnras/233.4.759}

\bibitem[{{Bellm} {et~al.}(2019){Bellm}, {Kulkarni}, {Graham}, {Dekany},
  {Smith}, {Riddle}, {Masci}, {Helou}, {Prince}, {Adams}, {Barbarino},
  {Barlow}, {Bauer}, {Beck}, {Belicki}, {Biswas}, {Blagorodnova}, {Bodewits},
  {Bolin}, {Brinnel}, {Brooke}, {Bue}, {Bulla}, {Burruss}, {Cenko}, {Chang},
  {Connolly}, {Coughlin}, {Cromer}, {Cunningham}, {De}, {Delacroix}, {Desai},
  {Duev}, {Eadie}, {Farnham}, {Feeney}, {Feindt}, {Flynn}, {Franckowiak},
  {Frederick}, {Fremling}, {Gal-Yam}, {Gezari}, {Giomi}, {Goldstein},
  {Golkhou}, {Goobar}, {Groom}, {Hacopians}, {Hale}, {Henning}, {Ho}, {Hover},
  {Howell}, {Hung}, {Huppenkothen}, {Imel}, {Ip}, {Ivezi{\'c}}, {Jackson},
  {Jones}, {Juric}, {Kasliwal}, {Kaspi}, {Kaye}, {Kelley}, {Kowalski},
  {Kramer}, {Kupfer}, {Landry}, {Laher}, {Lee}, {Lin}, {Lin}, {Lunnan},
  {Giomi}, {Mahabal}, {Mao}, {Miller}, {Monkewitz}, {Murphy}, {Ngeow},
  {Nordin}, {Nugent}, {Ofek}, {Patterson}, {Penprase}, {Porter}, {Rauch},
  {Rebbapragada}, {Reiley}, {Rigault}, {Rodriguez}, {van Roestel}, {Rusholme},
  {van Santen}, {Schulze}, {Shupe}, {Singer}, {Soumagnac}, {Stein}, {Surace},
  {Sollerman}, {Szkody}, {Taddia}, {Terek}, {Van Sistine}, {van Velzen},
  {Vestrand}, {Walters}, {Ward}, {Ye}, {Yu}, {Yan}, \& {Zolkower}}]{2019ZTF}
{Bellm}, E.~C., {Kulkarni}, S.~R., {Graham}, M.~J., {et~al.} 2019, \pasp, 131,
  018002, \dodoi{10.1088/1538-3873/aaecbe}

\bibitem[{{Bonnet-Bidaud} {et~al.}(1985){Bonnet-Bidaud}, {Motch}, \&
  {Mouchet}}]{1985A&A...143..313B}
{Bonnet-Bidaud}, J.~M., {Motch}, C., \& {Mouchet}, M. 1985, \aap, 143, 313

\bibitem[{Borucki {et~al.}(2010)Borucki, Koch, Basri, Batalha, Brown, Caldwell,
  Caldwell, Christensen-Dalsgaard, Cochran, DeVore,
  {et~al.}}]{borucki2010kepler}
Borucki, W.~J., Koch, D., Basri, G., {et~al.} 2010, Science, 327, 977

\bibitem[{{Boyd} {et~al.}(2017){Boyd}, {de Miguel}, {Patterson}, {Wood},
  {Barrett}, {Boardman}, {Brettman}, {Cejudo}, {Collins}, {Cook}, {Cook},
  {Foote}, {Fried}, {Gomez}, {Hambsch}, {Jones}, {Kemp}, {Koff}, {Koppelman},
  {Krajci}, {Lemay}, {Martin}, {McClusky}, {Menzies}, {Messier}, {Roberts},
  {Robertson}, {Rock}, {Sabo}, {Skillman}, {Ulowetz}, \&
  {Vanmunster}}]{Boyd2017}
{Boyd}, D.~R.~S., {de Miguel}, E., {Patterson}, J., {et~al.} 2017, \mnras, 466,
  3417, \dodoi{10.1093/mnras/stw3327}

\bibitem[{{Bruch}(2022)}]{Bruch2022}
{Bruch}, A. 2022, \mnras, 514, 4718, \dodoi{10.1093/mnras/stac1650}

\bibitem[{{Bruch}(2023{\natexlab{a}})}]{Bruch2023}
---. 2023{\natexlab{a}}, \mnras, 519, 352, \dodoi{10.1093/mnras/stac3493}

\bibitem[{{Bruch}(2023{\natexlab{b}})}]{BruchIII}
---. 2023{\natexlab{b}}, \mnras, \dodoi{10.1093/mnras/stad2089}

\bibitem[{{Chen} {et~al.}(2001){Chen}, {O'Donoghue}, {Stobie}, {Kilkenny}, \&
  {Warner}}]{2001MNRASChen}
{Chen}, A., {O'Donoghue}, D., {Stobie}, R.~S., {Kilkenny}, D., \& {Warner}, B.
  2001, \mnras, 325, 89, \dodoi{10.1046/j.1365-8711.2001.04322.x}

\bibitem[{Cleveland(1979)}]{cleveland1979robust}
Cleveland, W.~S. 1979, Journal of the American statistical association, 74, 829

\bibitem[{Court {et~al.}(2020)Court, Scaringi, Littlefield, Castro~Segura,
  Long, Maccarone, Altamirano, Degenaar, Wijnands, Shahbaz,
  {et~al.}}]{court2020ex}
Court, J., Scaringi, S., Littlefield, C., {et~al.} 2020, \mnras, 494, 4656,
  \dodoi{10.1093/mnras/staa1042}

\bibitem[{{D'Esterre}(1912)}]{1912Esterre}
{D'Esterre}, C.~R. 1912, AN, 192, 131

\bibitem[{{Drake} {et~al.}(2009){Drake}, {Djorgovski}, {Mahabal}, {Beshore},
  {Larson}, {Graham}, {Williams}, {Christensen}, {Catelan}, {Boattini},
  {Gibbs}, {Hill}, \& {Kowalski}}]{2009ApJ...696..870D}
{Drake}, A.~J., {Djorgovski}, S.~G., {Mahabal}, A., {et~al.} 2009, \apj, 696,
  870, \dodoi{10.1088/0004-637X/696/1/870}

\bibitem[{{Echevarr{\'\i}a} {et~al.}(1999){Echevarr{\'\i}a}, {Pineda}, \&
  {Costero}}]{1999RMxAA..35..135E}
{Echevarr{\'\i}a}, J., {Pineda}, L., \& {Costero}, R. 1999, \rmxaa, 35, 135

\bibitem[{{F{\"o}rster} {et~al.}(2021){F{\"o}rster}, {Cabrera-Vives},
  {Castillo-Navarrete}, {Est{\'e}vez}, {S{\'a}nchez-S{\'a}ez}, {Arredondo},
  {Bauer}, {Carrasco-Davis}, {Catelan}, {Elorrieta}, {Eyheramendy}, {Huijse},
  {Pignata}, {Reyes}, {Reyes}, {Rodr{\'\i}guez-Mancini}, {Ruz-Mieres},
  {Valenzuela}, {{\'A}lvarez-Maldonado}, {Astorga}, {Borissova}, {Clocchiatti},
  {De Cicco}, {Donoso-Oliva}, {Hern{\'a}ndez-Garc{\'\i}a}, {Graham},
  {Jord{\'a}n}, {Kurtev}, {Mahabal}, {Maureira}, {Mu{\~n}oz-Arancibia},
  {Molina-Ferreiro}, {Moya}, {Palma}, {P{\'e}rez-Carrasco}, {Protopapas},
  {Romero}, {Sabatini-Gacitua}, {S{\'a}nchez}, {San Mart{\'\i}n},
  {Sep{\'u}lveda-Cobo}, {Vera}, \& {Vergara}}]{2021AJ....161..242F}
{F{\"o}rster}, F., {Cabrera-Vives}, G., {Castillo-Navarrete}, E., {et~al.}
  2021, \aj, 161, 242, \dodoi{10.3847/1538-3881/abe9bc}

\bibitem[{{Gulbis} {et~al.}(2013){Gulbis}, {Kotze}, {Kotze}, {Worters},
  {Buckley}, {O'Donoghue}, \& {Shara}}]{2013Gulbis}
{Gulbis}, A.~A.~S., {Kotze}, M.~M., {Kotze}, E.~J., {et~al.} 2013, ATel, 5207,
  1

\bibitem[{{Hameury}(2020)}]{2020AdSpR..66.1004H}
{Hameury}, J.~M. 2020, Advances in Space Research, 66, 1004,
  \dodoi{10.1016/j.asr.2019.10.022}

\bibitem[{Harvey {et~al.}(1995)Harvey, Skillman, Patterson, \&
  Ringwald}]{harvey1995superhumps}
Harvey, D., Skillman, D.~R., Patterson, J., \& Ringwald, F. 1995, \pasp, 107,
  551, \dodoi{10.1086/133591}

\bibitem[{{Hou} {et~al.}(2020){Hou}, {Luo}, {Li}, \&
  {Qin}}]{2020AJ....159...43H}
{Hou}, W., {Luo}, A.~l., {Li}, Y.-B., \& {Qin}, L. 2020, \aj, 159, 43,
  \dodoi{10.3847/1538-3881/ab5962}

\bibitem[{Howell {et~al.}(2014)Howell, Sobeck, Haas, Still, Barclay, Mullally,
  Troeltzsch, Aigrain, Bryson, Caldwell, {et~al.}}]{howell2014k2}
Howell, S.~B., Sobeck, C., Haas, M., {et~al.} 2014, \pasp, 126, 398

\bibitem[{{I{\l}kiewicz} {et~al.}(2021){I{\l}kiewicz}, {Scaringi}, {Court},
  {Maccarone}, {Altamirano}, {Bradshaw}, {Degenaar}, {Fratta}, {Littlefield},
  {Shahbaz}, \& {Wijnands}}]{2021MNRAS.503.4050I}
{I{\l}kiewicz}, K., {Scaringi}, S., {Court}, J. M.~C., {et~al.} 2021, \mnras,
  503, 4050, \dodoi{10.1093/mnras/stab664}

\bibitem[{{Jayasinghe} {et~al.}(2018){Jayasinghe}, {Kochanek}, {Stanek},
  {Shappee}, {Holoien}, {Thompson}, {Prieto}, {Dong}, {Pawlak}, {Shields},
  {Pojmanski}, {Otero}, {Britt}, \& {Will}}]{2018MNRASJayasinghe}
{Jayasinghe}, T., {Kochanek}, C.~S., {Stanek}, K.~Z., {et~al.} 2018, \mnras,
  477, 3145, \dodoi{10.1093/mnras/sty838}

\bibitem[{Jenkins {et~al.}(2016)Jenkins, Twicken, \& McCauliff}]{Jenkins2016}
Jenkins, J.~M., Twicken, J., \& McCauliff, S. 2016, in Proc. SPIE, 9913,
  \dodoi{10.1117/12.2233418}

\bibitem[{{Kato} {et~al.}(2013){Kato}, {Hambsch}, \& {Maehara}}]{Kato2013PASJ}
{Kato}, T., {Hambsch}, F.-J., \& {Maehara}, e.~a. 2013, \pasj, 65, 23,
  \dodoi{10.1093/pasj/65.1.23}

\bibitem[{{Kato} \& {Hiroyuki}(2013)}]{2013PASJ...65...76K}
{Kato}, T., \& {Hiroyuki}, M. 2013, \pasj, 65, 76, \dodoi{10.1093/pasj/65.4.76}

\bibitem[{{Kato} {et~al.}(2009){Kato}, {Imada}, {Uemura}, {Nogami}, {Maehara},
  {Ishioka}, {Baba}, {Matsumoto}, {Iwamatsu}, {Kubota}, {Sugiyasu}, {Soejima},
  {Moritani}, {Ohshima}, {Ohashi}, {Tanaka}, {Sasada}, {Arai}, {Nakajima},
  {Kiyota}, {Tanabe}, {Imamura}, {Kunitomi}, {Kunihiro}, {Taguchi}, {Koizumi},
  {Yamada}, {Nishi}, {Kida}, {Tanaka}, {Ueoka}, {Yasui}, {Maruoka}, {Henden},
  {Oksanen}, {Moilanen}, {Tikkanen}, {Aho}, {Monard}, {Itoh}, {Dubovsky},
  {Kudzej}, {Dancikova}, {Vanmunster}, {Pietz}, {Bolt}, {Boyd}, {Nelson},
  {Krajci}, {Cook}, {Torii}, {Starkey}, {Shears}, {Jensen}, {Masi}, {Hynek},
  {Nov{\'a}k}, {Koci{\'a}n}, {Kr{\'a}l}, {Ku{\v{c}}{\'a}kov{\'a}}, {Kolasa},
  {{\v{S}}tastn{\'y}}, {Staels}, {Miller}, {Sano}, {Ponthi{\`e}re},
  {Miyashita}, {Crawford}, {Brady}, {Santallo}, {Richards}, {Martin},
  {Buczynski}, {Richmond}, {Kern}, {Davis}, {Crabtree}, {Beaulieu}, {Davis},
  {Aggleton}, {Morelle}, {Pavlenko}, {Andreev}, {Baklanov}, {Koppelman},
  {Billings}, {Urban{\v{c}}ok}, {{\"O}gmen}, {Heathcote}, {Gomez}, {Voloshina},
  {Retter}, {Mularczyk}, {Z{\l}oczewski}, {Olech}, {Kedzierski}, {Pickard},
  {Stockdale}, {Virtanen}, {Morikawa}, {Hambsch}, {Garradd}, {Gualdoni},
  {Geary}, {Omodaka}, {Sakai}, {Michel}, {C{\'a}rdenas}, {Gazeas}, {Niarchos},
  {Yushchenko}, {Mallia}, {Fiaschi}, {Good}, {Walker}, {James}, {Douzu},
  {Julian}, {Butterworth}, {Shugarov}, {Volkov}, {Chochol}, {Katysheva},
  {Rosenbush}, {Khramtsova}, {Kehusmaa}, {Reszelski}, {Bedient}, {Liller},
  {Pojma{\'n}ski}, {Simonsen}, {Stubbings}, {Schmeer}, {Muyllaert}, {Kinnunen},
  {Poyner}, {Ripero}, \& {Kriebel}}]{2009PASJ...61S.395K}
{Kato}, T., {Imada}, A., {Uemura}, M., {et~al.} 2009, \pasj, 61, S395,
  \dodoi{10.1093/pasj/61.sp2.S395}

\bibitem[{{Katz}(1973)}]{Katz1973NPhS..246...87K}
{Katz}, J.~I. 1973, Nature Physical Science, 246, 87,
  \dodoi{10.1038/physci246087a0}

\bibitem[{{Kimura} \& {Osaki}(2021)}]{2021PASJ...73.1225K}
{Kimura}, M., \& {Osaki}, Y. 2021, \pasj, 73, 1225,
  \dodoi{10.1093/pasj/psab069}

\bibitem[{{Kimura} {et~al.}(2020){Kimura}, {Osaki}, \&
  {Kato}}]{KimuraKIC94066522020PASJ}
{Kimura}, M., {Osaki}, Y., \& {Kato}, T. 2020, \pasj, 72, 94,
  \dodoi{10.1093/pasj/psaa088}

\bibitem[{{Kinemuchi} {et~al.}(2012){Kinemuchi}, {Barclay}, {Fanelli},
  {Pepper}, {Still}, \& {Howell}}]{2012PASP..124..963K}
{Kinemuchi}, K., {Barclay}, T., {Fanelli}, M., {et~al.} 2012, \pasp, 124, 963,
  \dodoi{10.1086/667603}

\bibitem[{{Kochanek} {et~al.}(2017){Kochanek}, {Shappee}, {Stanek}, {Holoien},
  {Thompson}, {Prieto}, {Dong}, {Shields}, {Will}, {Britt}, {Perzanowski}, \&
  {Pojma{\'n}ski}}]{2017PASPKochanek}
{Kochanek}, C.~S., {Shappee}, B.~J., {Stanek}, K.~Z., {et~al.} 2017, \pasp,
  129, 104502, \dodoi{10.1088/1538-3873/aa80d9}

\bibitem[{{Larwood}(1998)}]{1998MNRAS.299L..32L}
{Larwood}, J. 1998, \mnras, 299, L32, \dodoi{10.1046/j.1365-8711.1998.01978.x}

\bibitem[{Lasota(2001)}]{lasota2001disc}
Lasota, J.-P. 2001, New Astronomy Reviews, 45, 449,
  \dodoi{10.1016/S1387-6473(01)00112-9}

\bibitem[{{Lenz} \& {Breger}(2005)}]{Period04}
{Lenz}, P., \& {Breger}, M. 2005, Communications in Asteroseismology, 146, 53,
  \dodoi{10.1553/cia146s53}

\bibitem[{{Li} {et~al.}(2023){Li}, {Qian}, {Zhu}, {Liao}, {Zhao}, {Shi}, \&
  {Sun}}]{2023ApJS..266...28L}
{Li}, M.-Y., {Qian}, S.-B., {Zhu}, L.-Y., {et~al.} 2023, \apjs, 266, 28,
  \dodoi{10.3847/1538-4365/acca13}

\bibitem[{{Livio} \& {Pringle}(1994)}]{1994ApJ...427..956L}
{Livio}, M., \& {Pringle}, J.~E. 1994, Astrophysical Journal, 427, 956,
  \dodoi{10.1086/174202}

\bibitem[{{Montgomery}(2009)}]{Montgomery2009MNRAS}
{Montgomery}, M.~M. 2009, \mnras, 394, 1897,
  \dodoi{10.1111/j.1365-2966.2009.14487.x}

\bibitem[{{Motch}(1981)}]{Motch1981}
{Motch}, C. 1981, \aap, 100, 277

\bibitem[{{Ohshima} {et~al.}(2014){Ohshima}, {Kato}, \&
  {Pavlenko}}]{Ohshima2014PASJ}
{Ohshima}, T., {Kato}, T., \& {Pavlenko}, e.~a. 2014, \pasj, 66, 67,
  \dodoi{10.1093/pasj/psu038}

\bibitem[{{Ohshima} {et~al.}(2012){Ohshima}, {Kato}, {Pavlenko}, {Itoh}, {de
  Miguel}, {Krajci}, {Akazawa}, {Shiokawa}, {Stein}, {Baklanov}, {Samsonov},
  {Antonyuk}, {Andreev}, {Imamura}, {Hambsch}, {Maehara}, {Ruiz}, {Nakagawa},
  {Kasai}, {Boitnott}, {Virtanen}, \& {Miller}}]{Ohshima2012PASJ}
{Ohshima}, T., {Kato}, T., {Pavlenko}, E.~P., {et~al.} 2012, \pasj, 64, L3,
  \dodoi{10.1093/pasj/64.4.L3}

\bibitem[{{Olech} {et~al.}(2009){Olech}, {Rutkowski}, \&
  {Schwarzenberg-Czerny}}]{Olech2009MNRAS}
{Olech}, A., {Rutkowski}, A., \& {Schwarzenberg-Czerny}, A. 2009, \mnras, 399,
  465, \dodoi{10.1111/j.1365-2966.2009.15298.x}

\bibitem[{{Osaki}(1974)}]{1974PASJ...26..429O}
{Osaki}, Y. 1974, \pasj, 26, 429

\bibitem[{Osaki(1985)}]{osaki1985irradiation}
Osaki, Y. 1985, \aap, 144, 369

\bibitem[{{Osaki} \& {Kato}(2013)}]{Osaki2013b}
{Osaki}, Y., \& {Kato}, T. 2013, \pasj, 65, 95, \dodoi{10.1093/pasj/65.5.95}

\bibitem[{{Papaloizou} \& {Terquem}(1995)}]{1995MNRAS.274..987P}
{Papaloizou}, J. C.~B., \& {Terquem}, C. 1995, \mnras, 274, 987,
  \dodoi{10.1093/mnras/274.4.987}

\bibitem[{Patterson(1999)}]{patterson1999permanent}
Patterson, J. 1999, FRONTIERS SCIENCE SERIES, 61

\bibitem[{{Patterson} {et~al.}(1997){Patterson}, {Kemp}, {Saad}, {Skillman},
  {Harvey}, {Fried}, {Thorstensen}, \& {Ashley}}]{Patterson1997PASP}
{Patterson}, J., {Kemp}, J., {Saad}, J., {et~al.} 1997, \pasp, 109, 468,
  \dodoi{10.1086/133903}

\bibitem[{{Pavlenko} {et~al.}(2021){Pavlenko}, {Sosnovskii}, {Antonyuk},
  {Antonyuk}, {Pit'}, {Kokhirova}, {Rakhmatullaeva}, \&
  {Baklanov}}]{Pavlenko2021Ap}
{Pavlenko}, E.~P., {Sosnovskii}, A.~A., {Antonyuk}, K.~A., {et~al.} 2021,
  Astrophysics, 64, 293, \dodoi{10.1007/s10511-021-09690-3}

\bibitem[{{Peters} \& {Thorstensen}(2005)}]{2005PASPPeters}
{Peters}, C.~S., \& {Thorstensen}, J.~R. 2005, \pasp, 117, 1386,
  \dodoi{10.1086/497384}

\bibitem[{Polikar {et~al.}(1996)}]{polikar1996wavelet}
Polikar, R., {et~al.} 1996, The wavelet tutorial

\bibitem[{Pyrzas {et~al.}(2012)Pyrzas, G{\"a}nsicke, Thorstensen, Aungwerojwit,
  Boyd, Brady, Casares, Hickman, Marsh, Miller, {et~al.}}]{pyrzas2012hs}
Pyrzas, S., G{\"a}nsicke, B., Thorstensen, J., {et~al.} 2012, \pasp, 124, 204,
  \dodoi{10.1086/664959}

\bibitem[{Ramsay {et~al.}(2017)Ramsay, Wood, Cannizzo, Howell, \&
  Smale}]{ramsay2017v729}
Ramsay, G., Wood, M.~A., Cannizzo, J.~K., Howell, S.~B., \& Smale, A. 2017,
  \mnras, 469, 950, \dodoi{10.1093/mnras/stx859}

\bibitem[{Retter {et~al.}(2002)Retter, Chou, Bedding, \&
  Naylor}]{retter2002detection}
Retter, A., Chou, Y., Bedding, T., \& Naylor, T. 2002, \mnras, 330, L37

\bibitem[{Ricker {et~al.}(2015)Ricker, Winn, \& Vanderspek}]{Ricker2015journal}
Ricker, G., Winn, J., \& Vanderspek, R. 2015, JATIS, 1, 014003,
  \dodoi{10.1117/1.JATIS.1.1.014003}

\bibitem[{{Ringwald}(1995)}]{1995MNRASRingwald}
{Ringwald}, F.~A. 1995, \mnras, 274, 127, \dodoi{10.1093/mnras/274.1.127}

\bibitem[{{Smak}(2009)}]{smak2009origin}
{Smak}, J. 2009, \actaa, 59, 419, \dodoi{10.48550/arXiv.0910.2541}

\bibitem[{{Smak}(2013)}]{Smak2013}
---. 2013, \actaa, 63, 109, \dodoi{10.48550/arXiv.1301.0187}

\bibitem[{{Stefanov} \& {Stefanov}(2023)}]{2023MNRAS.520.3355S}
{Stefanov}, S.~Y., \& {Stefanov}, A.~K. 2023, \mnras, 520, 3355,
  \dodoi{10.1093/mnras/stad259}

\bibitem[{{Stobie} {et~al.}(1997){Stobie}, {Kilkenny}, {O'Donoghue}, {Chen},
  {Koen}, {Morgan}, {Barrow}, {Buckley}, {Cannon}, {Cass}, {Cranston},
  {Drinkwater}, {Hartley}, {Hawkins}, {Hughes}, {Humphries}, {MacGillivray},
  {McKenzie}, {Parker}, {Read}, {Russell}, {Savage}, {Thomson}, {Tritton},
  {Waldron}, {Warner}, \& {Watson}}]{1997MNRASStobie}
{Stobie}, R.~S., {Kilkenny}, D., {O'Donoghue}, D., {et~al.} 1997, \mnras, 287,
  848, \dodoi{10.1093/mnras/287.4.848}

\bibitem[{Sun {et~al.}(2023)Sun, Qian, \& Li}]{2023arXiv230905891S}
Sun, Q.-B., Qian, S.-B., \& Li, M.-Y. 2023, \apj, 955, 135,
  \dodoi{10.3847/1538-4357/ace183}

\bibitem[{{Sun} {et~al.}(2023{\natexlab{a}}){Sun}, {Qian}, {Zhu}, {Liao},
  {Zhao}, {Li}, {Shi}, \& {Li}}]{Sun2023arXiv}
{Sun}, Q.-B., {Qian}, S.-B., {Zhu}, L.-Y., {et~al.} 2023{\natexlab{a}}, \mnras,
  526, 3730, \dodoi{10.1093/mnras/stad1880}

\bibitem[{{Sun} {et~al.}(2022){Sun}, {Qian}, {Dong}, {Zhi}, {Han}, {Liu},
  {Chang}, {Liu}, {Xiang}, {Peng}, {Zhang}, {Zhang}, \& {Fern{\'a}ndez
  Laj{\'u}s}}]{sun2022study}
{Sun}, Q.-B., {Qian}, S.-B., {Dong}, A.-J., {et~al.} 2022, \na, 93, 101751,
  \dodoi{10.1016/j.newast.2021.101751}

\bibitem[{{Sun} {et~al.}(2023{\natexlab{b}}){Sun}, {Qian}, {Zhu}, {Dong},
  {Zhi}, {Liao}, {Zhao}, {Han}, {Liu}, {Zang}, {Li}, \& {Shi}}]{Sun2023MNRAS}
{Sun}, Q.-B., {Qian}, S.-B., {Zhu}, L.-Y., {et~al.} 2023{\natexlab{b}}, \mnras,
  518, 3901, \dodoi{10.1093/mnras/stac3272}

\bibitem[{{Sun} {et~al.}(2021){Sun}, {Cheng}, {Ye}, {Ding}, {Peng}, {Zhang},
  {Huo}, {Cui}, {Wang}, {Shi}, {Lin}, {Wu}, {Li}, {Feng}, {Yu}, {Ma}, {Li},
  {Liu}, {Zhang}, \& {Shao}}]{2021ApJS..257...65S}
{Sun}, Y., {Cheng}, Z., {Ye}, S., {et~al.} 2021, \apjs, 257, 65,
  \dodoi{10.3847/1538-4365/ac283a}

\bibitem[{{Szkody} \& {Mattei}(1984)}]{1984PASPSzkody}
{Szkody}, P., \& {Mattei}, J.~A. 1984, \pasp, 96, 988, \dodoi{10.1086/131464}

\bibitem[{{Tetarenko} {et~al.}(2018){Tetarenko}, {Lasota}, {Heinke}, {Dubus},
  \& {Sivakoff}}]{2018Natur.554...69T}
{Tetarenko}, B.~E., {Lasota}, J.~P., {Heinke}, C.~O., {Dubus}, G., \&
  {Sivakoff}, G.~R. 2018, Nature, 554, 69, \dodoi{10.1038/nature25159}

\bibitem[{{Twicken} {et~al.}(2010){Twicken}, {Chandrasekaran}, {Jenkins},
  {Gunter}, {Girouard}, \& {Klaus}}]{2010SPIE.7740E..1UT}
{Twicken}, J.~D., {Chandrasekaran}, H., {Jenkins}, J.~M., {et~al.} 2010, in
  Society of Photo-Optical Instrumentation Engineers (SPIE) Conference Series,
  Vol. 7740, Software and Cyberinfrastructure for Astronomy, ed. N.~M.
  {Radziwill} \& A.~{Bridger}, 77401U, \dodoi{10.1117/12.856798}

\bibitem[{Vogt(1982)}]{vogt1982z}
Vogt, N. 1982, \apj, 252, 653, \dodoi{10.1086/159592}

\bibitem[{Warner(1995)}]{warner1995cat}
Warner, B. 1995, Cambridge University Press, 28

\bibitem[{{Watson} {et~al.}(2006){Watson}, {Henden}, \&
  {Price}}]{2006SASSWatson}
{Watson}, C.~L., {Henden}, A.~A., \& {Price}, A. 2006, Society for Astronomical
  Sciences Annual Symposium, 25, 47

\bibitem[{{Wood} \& {Burke}(2007)}]{Wood2007ApJ...661.1042W}
{Wood}, M.~A., \& {Burke}, C.~J. 2007, Astrophysical Journal, 661, 1042,
  \dodoi{10.1086/516723}

\bibitem[{Wood {et~al.}(2011)Wood, Still, Howell, Cannizzo, \&
  Smale}]{wood2011v344}
Wood, M.~A., Still, M.~D., Howell, S.~B., Cannizzo, J.~K., \& Smale, A.~P.
  2011, \apj, 741, 105, \dodoi{10.1088/0004-637X/741/2/105}

\bibitem[{{Wood} {et~al.}(2011){Wood}, {Still}, {Howell}, {Cannizzo}, \&
  {Smale}}]{Wood2011ApJ}
{Wood}, M.~A., {Still}, M.~D., {Howell}, S.~B., {Cannizzo}, J.~K., \& {Smale},
  A.~P. 2011, \apj, 741, 105, \dodoi{10.1088/0004-637X/741/2/105}

\bibitem[{Wood {et~al.}(2009)Wood, Thomas, \& Simpson}]{wood2009sph}
Wood, M.~A., Thomas, D.~M., \& Simpson, J.~C. 2009, \mnras, 398, 2110,
  \dodoi{10.1111/j.1365-2966.2009.15252.x}

\end{thebibliography}
\bibliographystyle{aasjournal}
\end{document}